\documentclass[a4paper,times,3p,9pt,twocolumn]{elsarticle}
\usepackage{extsizes}

\usepackage{cuted}     %
\usepackage{caption}   %
\usepackage{float}     %

\usepackage{iftex}

\usepackage[dvipsnames]{xcolor}

\usepackage{mathtools}

\usepackage{amsmath}
\usepackage{amssymb}
\usepackage{amsthm}
\usepackage{siunitx}

\usepackage{makecell}

\usepackage{graphicx}
\usepackage{array}      %
\usepackage[figuresleft]{rotating}   %

\newsavebox{\refbox}

\usepackage{gensymb}

\ifxetex
    \usepackage[no-math]{fontspec}
    \usepackage[math-style=TeX,bold-style=TeX,mathbf=sym]{unicode-math}
\else

    \usepackage[utf8]{inputenc}
    
    \usepackage[T1]{fontenc}

    \usepackage{bm}

    \usepackage{pxfonts}
\fi

\usepackage{microtype}

\usepackage[x11names]{xcolor}

\usepackage{booktabs}
\usepackage{array}

\usepackage[hidelinks]{hyperref}

\usepackage{outlines}

\usepackage{graphicx}
\usepackage{pgf}
\usepackage{import}

\usepackage{rotating}

\usepackage{longtable}

\usepackage{algorithm}
\usepackage{algorithmic}

\RequirePackage{etoolbox}
\DeclarePairedDelimiterX\norm[1]{\lVert}{\rVert}{
\ifblank{#1}{\:\cdot\:}{#1}
}
\DeclarePairedDelimiterX\seminorm[1]{\lvert}{\rvert}{
\ifblank{#1}{\:\cdot\:}{#1}
}
\DeclarePairedDelimiterX\avg[1]{\langle}{\rangle}{
\ifblank{#1}{\:\cdot\:}{#1}
}
\DeclarePairedDelimiterX\ceil[1]{\lceil}{\rceil}{
\ifblank{#1}{\:\cdot\:}{#1}
}
\DeclarePairedDelimiterX\sprod[2]{(}{)}{
\ifblank{#1}{\cdot}{#1},\ifblank{#2}{\cdot}{#2}
}
\DeclarePairedDelimiterX\dprod[2]{\langle}{\rangle}{
\ifblank{#1}{\cdot}{#1},\ifblank{#2}{\cdot}{#2}
}

\DeclarePairedDelimiterX\set[2]{\{}{\}}
  {#1 \mathrel{}\mathclose{}\delimsize|\mathopen{}\mathrel{} #2}

\DeclareMathOperator{\optr}{tr}

\DeclareMathOperator{\opad}{ad}

\DeclareMathOperator{\opatantwo}{atan2}

\DeclareMathOperator{\arsinh}{arsinh}

\makeatletter
\renewcommand*\env@matrix[1][\arraystretch]{%
  \edef\arraystretch{#1}%
  \hskip -\arraycolsep
  \let\@ifnextchar\new@ifnextchar
  \array{*\c@MaxMatrixCols c}}
\makeatother

\definecolor{fireenginered}{rgb}{0.81, 0.09, 0.13}

\usepackage[font=footnotesize,labelfont=bf]{caption}
\usepackage[font=footnotesize,labelfont=bf]{subcaption}
\usepackage{multirow}

\usepackage{import}

\usepackage{tikz}
\usepackage{tikz-3dplot}
\usetikzlibrary{3d,calc}
\usetikzlibrary{arrows.meta,backgrounds,fit,positioning,petri}
\usetikzlibrary{shapes.geometric}
\usetikzlibrary{patterns}
\usetikzlibrary{fadings}

\usepackage[textsize=scriptsize]{todonotes}
\usepackage{setspace}

\theoremstyle{definition}
\newtheorem{rem}{Remark}

\def\statement{\begin{minipage}[t]{.75\textwidth}
       Preprint
       \end{minipage}}

\makeatletter
\def\ps@pprintTitle{%
     \let\@oddhead\@empty
     \let\@evenhead\@empty
     \def\@oddfoot{\footnotesize\itshape
       \statement\hfill\today}%
     \let\@evenfoot\@oddfoot}
\makeatother

\journal{}

\begin{document}
\begin{frontmatter}

\title{Giesekus Stick-Slip Singularity: Asymptotic Theory in the Log-Conformation Formulation}

\author{Florian Becker\corref{cor1}}
\ead{f.becker@dlr.de}
\author{Philipp Knechtges}
\ead{philipp.knechtges@dlr.de}
\cortext[cor1]{Corresponding author}
\address{German Aerospace Center (DLR), Institute of Software Technology, Department of High-Performance Computing, Cologne, Germany}

\begin{abstract}
In this paper, the log-conformation reformulation of a Giesekus fluid near a planar stick-slip singularity is investigated analytically. In what is, to our knowledge, the first use of the log-conformation formulation beyond its numerical purpose, we present results that match the asymptotic stress analysis of Evans [JNNFM 222 (2015) 24--33] for the stick, slip, and core regions under the assumption of a given Newtonian velocity field. Furthermore, the logarithmic reformulation allows us to extend the existing asymptotic theory substantially, showing, e.g., that the determinant of the conformation tensor is asymptotically constant, as well as thoroughly characterizing the transition behavior between the boundary layers and the core region. Combining these results, we obtain a composite asymptotic solution of the log-conformation equation that is valid uniformly throughout a close neighborhood of the singularity.
\end{abstract}

\begin{keyword}
Giesekus model\sep Log-conformation formulation\sep Stick-slip singularity\sep Asymptotic theory\sep Boundary layers
\end{keyword}
\end{frontmatter}

\section{Introduction}\label{sec:introduction}

Many applications in the polymer processing industry naturally involve free-surface flows and the transition from a confined flow, in which the polymer experiences drag from the walls of the confining vessel, into a region where these constraints are released more or less abruptly. For almost as long as polymers have been processed this way, e.g., in extrusion or coating applications, it has been known that free-surface disruptions such as sharkskin or melt rupture can occur~\cite{uhland1979anomale}, although the origin of these effects has not been conclusively settled to this day. Numerical approaches~\cite{varchanis2021origin} give strong indications that the transition from the stick to the slip regime, together with the general non-linearity of the constitutive models, is what leads to these bifurcation phenomena.

\begin{figure*}[h!]
    \centering
    \resizebox{0.75\textwidth}{!}{%
        \begin{tikzpicture}[>=stealth]

    \def\xmin{-6}
    \def\xmax{6}
    \def\H{4}
    \def\ubar{1}
    \def\stickstart{0}
    \def\wallthick{0.24}
    \def\labeloffset{0.3}

    \def\lw{1.6}          %
    \def\axislw{2.0}      %
    \def\axislen{1.5}     %
    \def\singrad{3.4}     %
    \def\txt{\large}      %

    \def\uin(#1){3*\ubar*(#1)/\H*(1-(#1)/(2*\H))}

    \definecolor{colorgrey}{HTML}{E4E4E4}
    
    \fill[colorgrey]
        (\xmax,0) --
        plot[domain=0:\H,samples=100,smooth] ({\xmax-\uin(\x)},{\x}) --
        (\xmax,\H) -- cycle;

    \fill[colorgrey]
        (\xmin,0) -- ({\xmin-\ubar},0) -- ({\xmin-\ubar},\H) -- (\xmin,\H) -- cycle;

    \fill[gray!35] (\stickstart,-\wallthick) rectangle (\xmax,0);
    \draw[
        pattern=north east lines,
        pattern color=black,
        draw=black,
        line width=\lw pt
    ] (\stickstart,-\wallthick) rectangle (\xmax,0);

    \fill[gray!35] (\xmin,-\wallthick) rectangle (\stickstart,0);
    \draw[draw=black, line width=\lw pt] (\xmin,-\wallthick) rectangle (\stickstart,0);

    \draw[gray,line width=\lw pt] (\xmax,0) -- ({\xmax-\uin(0)},0);
    \draw[gray,line width=\lw pt] (\xmax,\H) -- ({\xmax-\uin(\H)},\H);

    \draw[gray,line width=\lw pt,domain=0:\H,samples=100,smooth]
        plot ({\xmax-\uin(\x)},{\x});

    \foreach \yy in {1,2,3}{
        \draw[->,gray,line width=\lw pt] (\xmax,\yy) -- ({\xmax-\uin(\yy)},\yy);
    }

    \node[gray, font=\txt] at ({\xmax-\uin(2)-0.7},{2.3}) {$\vec{u}$};

    \draw[gray,line width=\lw pt] ({\xmin-\ubar},0) -- ({\xmin-\ubar},\H);
    \draw[gray,line width=\lw pt] (\xmin,0) -- ({\xmin-\ubar},0);
    \draw[gray,line width=\lw pt] (\xmin,\H) -- ({\xmin-\ubar},\H);

    \foreach \yy in {1,2,3}{
        \draw[->,gray,line width=\lw pt] (\xmin,\yy) -- ({\xmin-\ubar},\yy);
    }

    \def\smallradius{0.7}

    \definecolor{color1}{HTML}{7E76A5}
    \definecolor{color2}{HTML}{CBAAD0}
    
    \pgfmathsetmacro{\rsmall}{\smallradius}
    \pgfmathsetmacro{\thetacsmall}{2*asin(sqrt(2./3.))}
    
    \fill[color1]
        (0,0) -- (\rsmall,0)
        arc[start angle=0,end angle=\thetacsmall,radius=\rsmall] -- cycle;
    
    \fill[color2]
        (0,0) -- ({\rsmall*cos(\thetacsmall)},{\rsmall*sin(\thetacsmall)})
        arc[start angle=\thetacsmall,end angle=180,radius=\rsmall] -- cycle;
    
    \draw[dashed, line width=0.6*\lw pt] (-\rsmall,0)
        arc[start angle=180,end angle=0,radius=\rsmall];

    \draw[line width=\lw pt] (\xmin,0) -- (\stickstart,0);

    \draw[line width=\lw pt] (\xmin,0) -- (\xmin,\H);
    \draw[line width=\lw pt] (\xmax,0) -- (\xmax,\H);
    \draw[dashed,line width=\lw pt] (\xmin,\H) -- (\xmax,\H);

    \fill[black] (0,0) circle (\singrad pt);

    \draw[->,line width=\axislw pt] (0,0) -- ({\axislen},0)
        node[above, font=\txt] {$x$};
    \draw[->,line width=\axislw pt] (0,0) -- ({0},{\axislen})
        node[left, font=\txt] {$y$};

    \node[font=\txt, text depth=0pt] at ({(\xmin+\stickstart)/2},{-\wallthick-\labeloffset}) {slip};
    \node[font=\txt, text depth=0pt] at ({(\stickstart+\xmax)/2},{-\wallthick-\labeloffset}) {stick};
    \node[font=\txt] at ({(\xmin+\xmax)/2},{\H+\labeloffset}) {symmetry};

    \def\arrowmid{2}

\end{tikzpicture}
    }
    \caption{Illustration of the stick-slip benchmark in Cartesian coordinates, with the flow directed from right to left. Our upcoming asymptotic analysis considers a close neighborhood of the stick-slip singularity, marked by the purple half-disk.}
    \label{fig:stickslip_kartesian_illu}
\end{figure*}
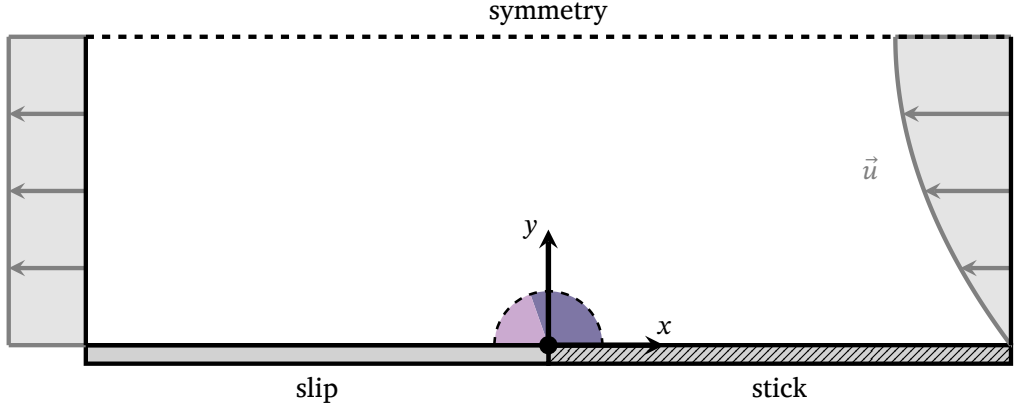

It is for this reason that the idealized situation of a fluid passing over a change of boundary conditions in the velocity field---from an upstream stick condition of zero flow at the wall, to a slip regime where the flow in the $x$-direction is unconstrained---has been studied not only for non-Newtonian fluids, but even in the Newtonian case~\cite{richardson1970stick}; the situation is visualized in Fig.~\ref{fig:stickslip_kartesian_illu}.

More precisely, we consider a planar Poiseuille flow with a parabolic velocity profile from right to left. At $x=0$ the boundary condition on the lower wall ($y=0$) switches from no-slip (stick, $x>0$) to shear-free (slip, $x<0$), giving rise to a stress singularity at the transition point $x=0$. A symmetry condition is imposed at the upper wall. Downstream of the transition, the flow relaxes toward a uniform plug profile.

As the constitutive model of our viscoelastic fluid, we will solely consider the Giesekus model~\cite{Giesekus1966, Giesekus1982} in this paper. The latter determines the symmetric conformation tensor $\mathbf{C}$, governed by
\begin{gather}\label{eqn:conformation_constitutive}
\begin{split}
	\partial_t \mathbf{C} + (\mathbf{u} \cdot \nabla) \mathbf{C}
		- \nabla\mathbf{u}\, \mathbf{C} - \mathbf{C}\, \nabla\mathbf{u}^T
		= - P(\mathbf{C})\, ,
\end{split}
\end{gather}
with velocity field $\mathbf{u} = (u_x, u_y)^T$ and velocity gradient field $\nabla \mathbf{u}$, where $[\nabla \mathbf{u}]_{ij}=\partial_j u_i$ is the Jacobian of $\mathbf{u}$. In general, $P(\mathbf{C})$ is a model-specific polynomial of the conformation tensor. For the Giesekus constitutive model it is specifically given by $P(\mathbf{C}) = \frac{1}{\lambda} \left(\mathbf{1} + \alpha \left(\mathbf{C} - \mathbf{1}\right)\right) \left(\mathbf{C} - \mathbf{1}\right)$, with relaxation time $\lambda>0$ and mobility parameter $\alpha\in(0,1)$.

Our main interest in this paper lies not only in locally characterizing the singularity of the polymeric conformation at the transition point (for a given velocity field), but also in quantifying the asymptotic solution as precisely as possible. Our main points of reference for the existing asymptotic results are~\cite{evans2015stick,evans2017stresses}, where it was already shown that, in the presence of a non-vanishing solvent viscosity, the velocity field can essentially be assumed to be Newtonian.

Even though our analysis is specifically concerned with the Giesekus stick-slip singularity, the work has to be understood in the broader context of the study of singularities of Upper-Convected Maxwell (UCM)-like fluids. On the one hand, the stick-slip transition is not the only configuration that produces a singularity in the velocity gradient $\nabla\mathbf{u}$; the reentrant corner is one particularly noteworthy example~\cite{hinch1993flow, rallison2004flow,evans2010re}. On the other hand, the UCM model relates to many viscoelastic models, including the Giesekus model, in the same way as the Euler equations relate to the Navier--Stokes equations as their high Reynolds number limit: the UCM model is the high Weissenberg number limit of the Giesekus model. It is thus natural to study the UCM fluid and its integration in the vicinity of singularities, as the singularities test the viscoelastic model in the high Weissenberg number limit. Most notable are the works by Renardy~\cite{Renardy199491, renardy1995matched, renardy1997high, renardy1997high2}. However, again comparable to the situation with the Euler equations~\cite{renardy2012boundary}, the UCM model is only a good approximation of the Giesekus model in advection-dominated regimes. Close to the boundary, where sharp and thin boundary layers may form, the non-linear terms of the respective viscoelastic constitutive equation dominate, which then determines the so-called viscometric behavior in this regime. For the Giesekus model, the work by Hagen and Renardy~\cite{hagen1997boundary} is most notable here.

Numerical investigations of the viscoelastic stick-slip problem date back at least to~\cite{owens1991spectral}, with more recent contributions in~\cite{evans2019numerical, evans2020testing,  wittschieber2023metric}. A common theme among numerical methods for viscoelastic fluids has been the use of the so-called log-conformation method~\cite{Fattal2004, Kwon2004, Hulsen2005, Damanik2010, Knechtges2014, Saramito2014, Knechtges2015, Knechtges2018, alves2021numerical, becker2023eigenvalue}. The latter replaces the conformation tensor $\mathbf{C}$ by its matrix logarithm $\mathbf{\Psi} = \log \mathbf{C}$ in order to mitigate the High Weissenberg Number Problem (HWNP) that had plagued computational rheology for decades~\cite{Keunings1986}. The advantages of the log-conformation approach are a) that the conformation tensor stays positive definite by design and b) that the exponential stress profiles are more palatable to numerical methods.

It is this last point that, to a large extent, also motivates the present paper: stress or strain singularities usually scale as $r^{-\mu}$, with $r$ being the distance to the singularity and $\mu$ some exponent which, as in our case, may also depend on the angle $\theta$ at which the singularity is approached. Studying the asymptotic behavior is thus mostly a question of determining the exponent and its angular dependence, making it, on the level of the stresses or the conformation field $\mathbf{C}$, a highly multiplicative endeavor. Using the logarithm, or the matrix logarithm, helps---as we will see---in mapping these multiplicative identifications into pure additions, thus giving us a tool with more quantitative control over the approximations made.

To our knowledge, this is thus the first time that the log-conformation approach is used outside of its numerical application to provide purely analytical insight into the mathematics of constitutive equations.

Hence, our exposition will start with a description of the log-conformation method based on the formulation developed in~\cite{becker2023eigenvalue,Knechtges2018}, which will then be reduced to the two-dimensional planar formulation of~\cite{Knechtges2014}. The advantage of the latter, in comparison to many descriptions of the log-conformation method in the literature, is that its eigenvalue-free form naturally facilitates our analysis.

After this short exposition of the log-conformation method and its formulation in polar coordinates, we derive an asymptotic solution of the constitutive equation at the stick-slip singularity by solving the equation first in the stick boundary layer, then in the core region, and finally in the slip boundary region; these partial solutions are lastly assembled into a full solution.

\section{Two-dimensional log-conformation formulation and its polar coordinate representation}\label{sec:log_conf_polar}

As indicated earlier, viscoelastic simulations based on constitutive models like Eq.~\eqref{eqn:conformation_constitutive} often suffer from the High Weissenberg Number Problem (HWNP) in flow regimes of high shear, where the numerical solution for the positive definite conformation tensor $\mathbf{C}$ locally loses this property and simulations tend to break down. The log-conformation (log-conf) formulation was conceived specifically to solve this problem: instead of solving Eq.~\eqref{eqn:conformation_constitutive} directly, the conformation tensor $\mathbf{C}$ is replaced by its matrix logarithm $\mathbf{\Psi}$, which is governed by the log-conf equation, the logarithmic counterpart of the constitutive Eq.~\eqref{eqn:conformation_constitutive}. The conformation tensor is recovered via $\mathbf{C} = \exp\mathbf{\Psi}$, which by design ensures positive definiteness of the numerical result and significantly improves the stability of the corresponding simulations. Although we do not perform any numerical simulations of partial differential equations in this paper, the log-conf equation still plays a central role in this work, and in our notation it reads
\begin{gather}
\begin{split}
	\partial_t \mathbf{\Psi} + (\mathbf{u} \cdot \nabla) \mathbf{\Psi}
		=& -\mathbf{\Psi} \omega(\mathbf{u}) + \omega(\mathbf{u}) \mathbf{\Psi} \\
		&\quad + 2\, f(\opad \mathbf{\Psi})\, \epsilon(\mathbf{u})
		- P(e^{\mathbf{\Psi}}) e^{-\mathbf{\Psi}}\, ,
\end{split}
\label{eqn:logconf}
\end{gather}
with the symmetric log-conf tensor $\mathbf{\Psi} = \log \mathbf{C}$, the velocity field $\mathbf{u}$, the vorticity tensor $\omega(\mathbf{u}) \coloneqq (\nabla \mathbf{u} - \nabla \mathbf{u}^T)/2$, and the strain tensor $\epsilon(\mathbf{u}) \coloneqq (\nabla \mathbf{u} + \nabla \mathbf{u}^T)/2$. The function $f$ is given here by
\begin{gather}
    f(x) = \frac{x/2}{\tanh(x/2)}\, ,
\end{gather}
where
\begin{gather*}
    \opad \mathbf{A}\, (\mathbf{B}) \coloneqq [\mathbf{A}, \mathbf{B}]
\end{gather*}
is the linear operator that maps any matrix $\mathbf{B}$ to the commutator $[\mathbf{A},\mathbf{B}] = \mathbf{A}\mathbf{B} - \mathbf{B}\mathbf{A}$, with $\mathbf{A},\mathbf{B} \in \mathbb{R}^{d\times d}$. To relate this to many of the common log-conf formulations, note that the eigenvalue decomposition $\mathbf{\Psi} = \sum_i \lambda_i \mathbf{P}_i$, with eigenvalues $\lambda_i$ and the associated spectral projectors $\mathbf{P}_i$, yields
\begin{gather*}
    \opad \mathbf{\Psi}\, \epsilon(\mathbf{u}) = \sum_{i,j} (\lambda_i - \lambda_j) \mathbf{P}_i \epsilon(\mathbf{u}) \mathbf{P}_j \, .
\end{gather*}
Analogously, for any power of $\opad  \mathbf{\Psi}$, and thus for any polynomial $p(x)$, one can show
\begin{gather*}
    p(\opad \mathbf{\Psi})\, \epsilon(\mathbf{u}) = \sum_{i,j} p(\lambda_i - \lambda_j) \mathbf{P}_i \epsilon(\mathbf{u}) \mathbf{P}_j \, .
\end{gather*}
Using the Stone--Weierstraß theorem, this allows us to evaluate $f(\opad \mathbf{\Psi})\, \epsilon(\mathbf{u})$ via the identity
\begin{align}\label{eqn:firsteigenvallogconf}
    f(\opad \mathbf{\Psi})\, \epsilon(\mathbf{u})
        &= \sum_{i,j} f(\lambda_i - \lambda_j) \mathbf{P}_i \epsilon(\mathbf{u}) \mathbf{P}_j\, ,
\end{align}
which, modulo the aggregation of terms, is the eigenvalue-based evaluation of this term as found in many implementations; cf.~\cite{becker2023eigenvalue} for more details.

In the following, we are solely interested in the two-dimensional planar situation, i.e., by design there can be at most two distinct eigenvalues. In order to further analyze the resulting system of equations and its individual components, we denote the diagonal entries by ${\Psi}_{11}$ and ${\Psi}_{22}$ and the off-diagonal entry by ${\Psi}_{12}$, and define the half-difference of the diagonal entries
\begin{align}
    \gamma(\mathbf{\Psi}) = \tfrac{1}{2}(\Psi_{11} - \Psi_{22})\, ,
\end{align}
together with the (quarter) discriminant
\begin{align}
    D = \gamma(\mathbf{\Psi})^2 + \Psi_{12}^2\, ,
\end{align}
as well as the orientation angle
\begin{align}
    \beta = \opatantwo(\Psi_{12}, \gamma(\mathbf{\Psi}))\, .
\end{align}

As can be verified by matrix multiplication and elementary trigonometry, the eigenvalue decomposition of $\mathbf{\Psi}$ can then be readily expressed in these new variables as
\begin{align}\label{eqn:eigvaldecomppsi}
    \mathbf{\Psi} =& \mathbf{Q}\begin{pmatrix} \frac{\optr\mathbf{\Psi}}{2} + \sqrt{D} & \\
            & \frac{\optr{\mathbf{\Psi}}}{2} - \sqrt{D}\end{pmatrix}\mathbf{Q}^T\\&\qquad\mbox{with }
            \mathbf{Q} = \begin{pmatrix}[1.8] \cos\frac{\beta}{2} & - \sin\frac{\beta}{2}\\
                \sin\frac{\beta}{2} & \cos\frac{\beta}{2}\end{pmatrix}\, ,\nonumber
\end{align}
and is therefore fully determined by $\optr \mathbf{\Psi} = \Psi_{11} + \Psi_{22}$, $\sqrt{D}$, and $\beta$. Consequently, instead of working with the tensor components, our subsequent analysis focuses on the triplet $(\optr\mathbf{\Psi}, \sqrt{D}, \beta)$, whose entries carry distinct geometric and physical meaning:
\begin{itemize}
    \item $\operatorname{tr}\mathbf{\Psi}$ is the isotropic part of $\mathbf{\Psi}$ and equals $\log \det \mathbf{C}$, i.e., the logarithmic volume change of the conformation tensor;
    \item $\sqrt{D}$ is the magnitude of the deviatoric part of $\mathbf{\Psi}$ and equals half the eigenvalue difference, quantifying the degree of stress anisotropy;
    \item $\beta/2$ encodes the orientation of the principal axes.
\end{itemize}

The matrix exponential then admits the closed-form expressions
\begin{align}\label{eqn:matrixexp2d}
    e^{\mathbf{\Psi}} =&\; e^{\operatorname{tr} \mathbf{\Psi} / 2}
        \begin{pmatrix}
            \gamma(\mathbf{\Psi}) \frac{\sinh \sqrt{D}}{\sqrt{D}} + \cosh\sqrt{D} &
            \Psi_{12} \frac{\sinh\sqrt{D}}{\sqrt{D}} \\
            \mbox{sym.} &
            -\gamma(\mathbf{\Psi}) \frac{\sinh \sqrt{D}}{\sqrt{D}} + \cosh\sqrt{D}
        \end{pmatrix} \\
    e^{-\mathbf{\Psi}} =&\; e^{-\operatorname{tr} \mathbf{\Psi} / 2}
        \begin{pmatrix}
            -\gamma(\mathbf{\Psi}) \frac{\sinh \sqrt{D}}{\sqrt{D}} + \cosh\sqrt{D} &
            -\Psi_{12} \frac{\sinh\sqrt{D}}{\sqrt{D}} \\
            \mbox{sym.} &
            \gamma(\mathbf{\Psi}) \frac{\sinh \sqrt{D}}{\sqrt{D}} + \cosh\sqrt{D}
        \end{pmatrix} \, ,\label{eqn:invmatrixexp2d}
\end{align}
which allow $e^{\pm\mathbf{\Psi}}$ to be evaluated without any series truncation.

For the further analysis of $f(\opad \mathbf{\Psi})\, \epsilon(\mathbf{u})$ based on Eq.~\eqref{eqn:firsteigenvallogconf}, we introduce the eigenvectors $\mathbf{e}_1 = (\cos\beta/2,\sin\beta/2)^T$ and $\mathbf{e}_2 = (-\sin\beta/2,\cos\beta/2)^T$, with the spectral projectors given by $\mathbf{P}_i = \mathbf{e}_i\mathbf{e}_i^T$.

First, noting that $f(0) = 1$, we split off this part and thus obtain
\begin{align*}
    f(\opad \mathbf{\Psi})\, \epsilon(\mathbf{u})
        &= \epsilon(\mathbf{u}) + \sum_{i,j} (f(\lambda_i - \lambda_j)-1) \mathbf{P}_i \epsilon(\mathbf{u}) \mathbf{P}_j\, ,
\end{align*}
which then, noting that $f$ is an even function, further simplifies to
\begin{align*}
    f(\opad \mathbf{\Psi})\, \epsilon(\mathbf{u})
        &= \epsilon(\mathbf{u}) + (f(\lambda_1 - \lambda_2)-1) \left(\mathbf{P}_1 \epsilon(\mathbf{u}) \mathbf{P}_2 + \mathbf{P}_2 \epsilon(\mathbf{u}) \mathbf{P}_1\right)\\
        &= \epsilon(\mathbf{u}) + h(D)\, \sqrt{D} \mathbf{e}_1^T \epsilon(\mathbf{u}) \mathbf{e}_2\; \sqrt{D} \left(\mathbf{e}_1\mathbf{e}_2^T + \mathbf{e}_2\mathbf{e}_1^T\right)\, ,
\end{align*}
where the function $h$ is given by
\begin{align}\label{eqn:hfunc}
    h(x) =& \frac{1}{x}\left(f\left(2\sqrt{x}\right)-1\right)\nonumber\\
        =& \frac{1}{x} \left(\frac{\sqrt{x}}{\tanh \sqrt{x}} - 1\right)\, .
\end{align}
Direct calculations then show that
\begin{align*}
    \sqrt{D} \mathbf{e}_1^T \epsilon(\mathbf{u}) \mathbf{e}_2
        =& \gamma(\mathbf{\Psi})\epsilon_{12} - \Psi_{12} \gamma(\epsilon(\mathbf{u}))\\
    \sqrt{D} \left(\mathbf{e}_1\mathbf{e}_2^T + \mathbf{e}_2\mathbf{e}_1^T\right)
        =& \begin{pmatrix} -\Psi_{12} & \gamma(\mathbf{\Psi}) \\
                    \gamma(\mathbf{\Psi}) & \Psi_{12}\end{pmatrix}\, .
\end{align*}
In total, we have rederived the representation of $f(\opad \mathbf{\Psi})\, \epsilon(\mathbf{u})$ as first given in~\cite[Theorem~2]{Knechtges2014}:
\begin{align}\label{eqn:logconf-paper2014}
\begin{split}
    f(\opad \mathbf{\Psi})\, \epsilon(\mathbf{u})
        =&\; \epsilon(\mathbf{u}) + \begin{pmatrix} -\Psi_{12} & \gamma(\mathbf{\Psi}) \\
                    \gamma(\mathbf{\Psi}) & \Psi_{12}\end{pmatrix} \\ 
        &\qquad\quad \cdot \left[\gamma(\mathbf{\Psi})\epsilon_{12} - \Psi_{12} \gamma(\epsilon(\mathbf{u}))\right] h(D)\, .
\end{split}
\end{align}

For the subsequent analysis, it is more convenient to consider the log-conf equation in polar coordinates.
Introducing the rotation matrix $\mathbf{O} = \left[\begin{matrix} \mathbf{\hat{e}_r} & \mathbf{\hat{e}_\theta}\end{matrix}\right]$ collecting the orthonormal polar basis vectors, we obtain
\begin{align*}
    \mathbf{O} =& \begin{pmatrix} \cos\theta & -\sin \theta\\ \sin\theta & \cos\theta \end{pmatrix}\, .
\end{align*}
The log-conformation tensor in the rotated reference frame is then given by $\mathbf{\tilde{\Psi}} = \mathbf{O}^T \mathbf{\Psi} \mathbf{O}$.

Given this transformation, note that $\optr\mathbf{\tilde{\Psi}}=\optr\mathbf{\Psi}$ and $\sqrt{D}$ are invariant under this change of reference frame. Regarding the transformation law of $\beta$, note that
\begin{align}\label{eqn:rot_identity_beta}
    \mathbf{O}^T \mathbf{Q} =& \begin{pmatrix} \cos(\beta/2 - \theta) & - \sin(\beta/2 -\theta)\\
            \sin(\beta/2 -\theta) & \cos(\beta/2 - \theta)\end{pmatrix}\, ,
\end{align}
thus motivating the definition of $\tilde{\beta} = \beta - 2\theta$ as the polar counterpart of $\beta$, such that the eigenvalue decomposition in~\eqref{eqn:eigvaldecomppsi} still holds true when $\mathbf{\Psi}$ and $\beta$ are replaced by their polar counterparts.

Note that most of the considerations above regarding the representations of $e^{\mathbf{\Psi}}$ and $f(\opad\mathbf{\Psi})$ have been agnostic to the coordinate frame and are thus valid in polar coordinates as well. This also holds for almost all other terms in the log-conformation equation~\eqref{eqn:logconf}, as they transform transparently under the rotation to polar coordinates; e.g., by inserting $\mathbf{O}\,\mathbf{O}^T = \mathbf{I}$ twice, one obtains $\mathbf{O}^T[\mathbf{\Psi},\,\omega(\mathbf{u})]\,\mathbf{O} = [\mathbf{O}^T\mathbf{\Psi}\,\mathbf{O},\, \mathbf{O}^T\omega(\mathbf{u})\,\mathbf{O}] = [\tilde{\mathbf{\Psi}},\,\tilde{\omega}(\mathbf{u})]$.

The only exception is the advective derivative, which needs to be addressed separately. In particular,
\begin{align*}
    \mathbf{O}^T \partial_\theta \mathbf{O}
        =& \begin{pmatrix} \cos\theta & \sin \theta\\ -\sin\theta & \cos\theta \end{pmatrix}
            \cdot \begin{pmatrix} -\sin\theta & -\cos \theta\\ \cos\theta & -\sin\theta \end{pmatrix}\\
        =& \begin{pmatrix} 0 & -1\\ 1 & 0 \end{pmatrix} \coloneqq \mathbf{A} = - \mathbf{A}^T \, , 
\end{align*}
and we recall that $\mathbf{\tilde{\Psi}} = \mathbf{O}^T \mathbf{\Psi} \mathbf{O}$ is the matrix representation of $\mathbf{\Psi}$ in the polar frame and denote by $(u_r, u_\theta)^T = \mathbf{O}^T \mathbf{u} = \mathbf{O}^T (u_x, u_y)^T$ the corresponding velocity components. Since $\mathbf{u} \cdot \nabla = u_x\partial_x + u_y\partial_y = u_r\partial_r + \tfrac{u_\theta}{r}\partial_\theta$ and $\mathbf{O}$ is $r$-independent, the polar reformulation of the advective term in the Giesekus log-conf equation becomes
\begin{align*}
    \mathbf{O}^T\left[(\mathbf{u} \cdot \nabla)\mathbf{\Psi}\right]\mathbf{O}
    &= u_r\,\mathbf{O}^T(\partial_r \mathbf{\Psi})\,\mathbf{O}
        + \frac{u_\theta}{r}\,\mathbf{O}^T(\partial_\theta \mathbf{\Psi})\,\mathbf{O}\\
    &= u_r \partial_r \mathbf{\tilde{\Psi}}
        + \frac{u_\theta}{r}\Big(\partial_\theta \mathbf{\tilde{\Psi}}
            - (\partial_\theta \mathbf{O}^T)\,\mathbf{\Psi}\,\mathbf{O}
            - \mathbf{O}^T\mathbf{\Psi}\,(\partial_\theta \mathbf{O})\Big)\\
    &= \left(u_r \partial_r + \frac{u_\theta}{r} \partial_\theta\right)\mathbf{\tilde{\Psi}}
        - \frac{u_\theta}{r}\left(\mathbf{A}^T\mathbf{\tilde{\Psi}}
            + \mathbf{\tilde{\Psi}}\,\mathbf{A}\right)\\
    &= \left(u_r \partial_r + \frac{u_\theta}{r} \partial_\theta\right)\mathbf{\tilde{\Psi}}
        - \frac{u_\theta}{r}\,[\mathbf{\tilde{\Psi}},\, \mathbf{A}]\, .
\end{align*}

In total, we thus obtain the following constitutive log-conformation equation in polar coordinates:
\begin{gather}
\begin{split}
    \partial_t \mathbf{\tilde{\Psi}} + \left(u_r \partial_r + \frac{u_\theta}{r} \partial_\theta\right) \mathbf{\tilde{\Psi}} 
        =&\; [\mathbf{\tilde{\Psi}}, \frac{u_\theta}{r} \mathbf{A} - \tilde{\omega}(\mathbf{u})] \\
        &\quad + 2 f(\opad \mathbf{\tilde{\Psi}}) \tilde{\epsilon}(\mathbf{u}) - P(e^{\mathbf{\tilde{\Psi}}})e^{-\mathbf{\tilde{\Psi}}}\, .
\end{split}
\label{eqn:polar_logconf}
\end{gather}
In particular, we can use the expressions for $f(\opad \mathbf{\tilde{\Psi}})\, \tilde{\epsilon}(\mathbf{u})$ and $e^{\pm\mathbf{\tilde{\Psi}}}$ as derived above in~\eqref{eqn:logconf-paper2014} and in~\eqref{eqn:matrixexp2d}, \eqref{eqn:invmatrixexp2d}, respectively.

\section{Stick-slip problem formulation}

For our asymptotic study, only a small neighborhood of the singularity is of interest, which we consider in polar coordinates. Assuming that the Newtonian stresses dominate in the vicinity of the singularity, the singular behavior of the velocity field and the streamlines is, up to a constant $C$, fully determined by the Newtonian part. We furthermore consider the steady-state regime, so that the time derivative $\partial_t \mathbf{\tilde{\Psi}}$ is dropped from Eq.~\eqref{eqn:polar_logconf}.

\begin{figure*}[h!]
    \centering
    \resizebox{\textwidth}{!}{%
        \begin{tikzpicture}[>=Latex]

\def\lw{1.2}
\def\txt{\normalsize}
\def\stxt{\small}
\def\wallthick{0.24}
\def\labeloffset{0.3}
\def\singrad{3.4}
\def\cartlw{1.0}
\def\cartlen{0.85}     %

\definecolor{color1}{HTML}{7E76A5}
\definecolor{color2}{HTML}{CBAAD0}

\def\R{4}              %
\def\lnrspan{6}        %
\def\panelgap{4.2}     %
\def\arrowclear{1.3}   %

\def\Rtheta{1.5}
\def\rAxisExtend{0.5}
\def\thetaarrow{25}

\pgfmathsetmacro{\thetac}{2*asin(sqrt(2./3.))}
\pgfmathsetmacro{\thetacrad}{\thetac*pi/180.0}

\pgfmathsetmacro{\H}{\R}
\pgfmathsetmacro{\Sy}{\H/pi}            %
\pgfmathsetmacro{\W}{\lnrspan*\Sy}
\pgfmathsetmacro{\thetacY}{\thetacrad*\Sy}
\pgfmathsetmacro{\shiftx}{\R+\panelgap}

\pgfmathsetmacro{\thA}{\thetac/4}
\pgfmathsetmacro{\thB}{\thetac/2}
\pgfmathsetmacro{\thC}{3*\thetac/4}
\pgfmathsetmacro{\thAout}{173.7474395447096}
\pgfmathsetmacro{\thBout}{157.49908401545932}
\pgfmathsetmacro{\thCout}{135.30459670218917}
\pgfmathsetmacro{\KA}{\R*(sin(\thA/2))^(4./3.)*(cos(\thA/2))^(2./3.)}
\pgfmathsetmacro{\KB}{\R*(sin(\thB/2))^(4./3.)*(cos(\thB/2))^(2./3.)}
\pgfmathsetmacro{\KC}{\R*(sin(\thC/2))^(4./3.)*(cos(\thC/2))^(2./3.)}

\def\rstream(#1,#2){(#1)/((sin((#2)/2))^(4./3.)*(cos((#2)/2))^(2./3.))}

\def\xstream(#1,#2){
    (4./3.)*ln(sin((#1)/2)/sin((#2)/2))
    + (2./3.)*ln(cos((#1)/2)/cos((#2)/2))
}

\begin{scope}
    \pgfmathsetmacro{\xc}{\R*cos(\thetac)}
    \pgfmathsetmacro{\yc}{\R*sin(\thetac)}
    \pgfmathsetmacro{\rAxisEndX}{(\Rtheta+\rAxisExtend)*cos(\thetaarrow)}
    \pgfmathsetmacro{\rAxisEndY}{(\Rtheta+\rAxisExtend)*sin(\thetaarrow)}
    \pgfmathsetmacro{\thetalabelangle}{0.5*\thetaarrow}
    \pgfmathsetmacro{\thetalabelx}{(\Rtheta+0.35)*cos(\thetalabelangle)}
    \pgfmathsetmacro{\thetalabely}{(\Rtheta-0.35)*sin(\thetalabelangle)}

    \fill[color1] (0,0) -- (\R,0)
        arc[start angle=0,end angle=\thetac,radius=\R] -- cycle;
    \fill[color2] (0,0) -- (\xc,\yc)
        arc[start angle=\thetac,end angle=180,radius=\R] -- cycle;

    \draw[line width=\lw pt] (-\R,0) arc[start angle=180,end angle=0,radius=\R];

    \fill[gray!35] (-\R,-\wallthick) rectangle (0,0);
    \draw[draw=black, line width=\lw pt] (-\R,-\wallthick) rectangle (0,0);

    \fill[gray!35] (0,-\wallthick) rectangle (\R,0);
    \draw[pattern=north east lines, pattern color=black, draw=black, line width=\lw pt]
          (0,-\wallthick) rectangle (\R,0);

    \draw[->, black, line width=0.9*\lw pt, smooth, variable=\t,
          domain=\thA:\thAout, samples=200]
        plot ({\rstream(\KA,\t)*cos(\t)},{\rstream(\KA,\t)*sin(\t)});
    \draw[->, black, line width=0.9*\lw pt, smooth, variable=\t,
          domain=\thB:\thBout, samples=200]
        plot ({\rstream(\KB,\t)*cos(\t)},{\rstream(\KB,\t)*sin(\t)});
    \draw[->, black, line width=0.9*\lw pt, smooth, variable=\t,
          domain=\thC:\thCout, samples=200]
        plot ({\rstream(\KC,\t)*cos(\t)},{\rstream(\KC,\t)*sin(\t)});

    \draw[->, black, line width=\lw pt] (0,0) -- (\rAxisEndX,\rAxisEndY);
    \node[font=\txt, anchor=south west] at (\rAxisEndX - 0.1,\rAxisEndY + 0.05) {$r$};
    \draw[->, black, line width=\lw pt]
        (\Rtheta,0) arc[start angle=0,end angle=\thetaarrow,radius=\Rtheta];
    \node[font=\txt] at (\thetalabelx,\thetalabely) {$\theta$};

    \draw[->, black, line width=\cartlw pt] (0,0) -- (\cartlen,0);
    \draw[->, black, line width=\cartlw pt] (0,0) -- (0,\cartlen);
    \node[font=\stxt, anchor=south west, inner sep=2pt] at ({\cartlen-0.04},0.04) {$x$};
    \node[font=\stxt, anchor=south west, inner sep=1pt] at (0.1,{\cartlen-0.10}) {$y$};

    \fill[black] (0,0) circle (\singrad pt);

    \node[font=\txt, text depth=0pt, anchor=east] at ({-\R-0.15},0) {$\theta=\pi$};
    \node[font=\txt, text depth=0pt, anchor=west] at ({ \R+0.15},0) {$\theta=0$};
    \node[font=\txt, text depth=0pt] at ({-\R/2},{-\wallthick-\labeloffset}) {slip};
    \node[font=\txt, text depth=0pt] at ({ \R/2},{-\wallthick-\labeloffset}) {stick};
    \node[above left, font=\txt, align=right] at (\xc,\yc) {$\theta=\theta_c$};
    \node[font=\txt, rotate={\thetac+180}, anchor=south]
        at ({0.5*\xc + 0.02*sin(\thetac)},{0.5*\yc - 0.02*cos(\thetac)}) {core};
    \node[font=\txt, black, text depth=0pt] at ({\R/2-1.0},{1.7}) {$\vec{\mathbf{u}}$};
\end{scope}

\pgfmathsetmacro{\arrY}{\H/2}
\pgfmathsetmacro{\axL}{\R+\arrowclear}
\pgfmathsetmacro{\axR}{\shiftx-\arrowclear}
\draw[->, black, line width=1.4pt] (\axL,\arrY) -- (\axR,\arrY);
\node[font=\stxt, anchor=south] at ({(\axL+\axR)/2},{\arrY+0.14})
    {$(x,y)\;\longmapsto\;(\ln r,\theta)$};

\begin{scope}[shift={(\shiftx,0)}]

    \fill[color1] (0,0) rectangle (\W,\thetacY);
    \fill[color2] (0,\thetacY) rectangle (\W,\H);

    \fill[gray!35] (0,-\wallthick) rectangle (\W,0);
    \draw[pattern=north east lines, pattern color=black, draw=black, line width=\lw pt]
          (0,-\wallthick) rectangle (\W,0);

    \fill[gray!35] (0,\H) rectangle (\W,{\H+\wallthick});
    \draw[draw=black, line width=\lw pt] (0,\H) rectangle (\W,{\H+\wallthick});

    \draw[line width=\lw pt] (\W,0) -- (\W,\H);

    \draw[->, black, line width=0.9*\lw pt, smooth, variable=\t,
          domain=\thA:\thAout, samples=200]
        plot ({\W + (\xstream(\thA,\t))*\Sy},{\t*pi/180.0*\Sy});
    \draw[->, black, line width=0.9*\lw pt, smooth, variable=\t,
          domain=\thB:\thBout, samples=200]
        plot ({\W + (\xstream(\thB,\t))*\Sy},{\t*pi/180.0*\Sy});
    \draw[->, black, line width=0.9*\lw pt, smooth, variable=\t,
          domain=\thC:\thCout, samples=200]
        plot ({\W + (\xstream(\thC,\t))*\Sy},{\t*pi/180.0*\Sy});

    \draw[black, line width=\lw pt] (0,0) -- (0,\H);

    \draw[->, black, line width=\cartlw pt] (0,0) -- (\cartlen,0);
    \draw[->, black, line width=\cartlw pt] (0,0) -- (0,\cartlen);
    \node[font=\stxt, anchor=south west, inner sep=2pt] at ({\cartlen-0.04},0.04) {$\ln r$};
    \node[font=\stxt, anchor=south west, inner sep=1pt] at (0.1,{\cartlen-0.10}) {$\theta$};

    \node[font=\txt, text depth=0pt] at ({\W/2},{-\wallthick-\labeloffset}) {stick};
    \node[font=\txt, text depth=0pt] at ({\W/2},{\H+\wallthick+\labeloffset}) {slip};
    \node[font=\txt] at ({\W/2},{\thetacY+\labeloffset}) {core};
    \node[font=\txt, rotate=90, text depth=0pt] at ({-\labeloffset},{\H/2}) {singularity};

    \node[font=\txt, black, text depth=0pt] at ({\W-1.2*\Sy},{0.8*\Sy}) {$\vec{\mathbf{u}}$};

    \node[font=\txt, anchor=west] at ({\W+0.2},0)        {$\theta=0$};
    \node[font=\txt, anchor=west] at ({\W+0.2},\thetacY) {$\theta=\theta_c$};
    \node[font=\txt, anchor=west] at ({\W+0.2},\H)       {$\theta=\pi$};
\end{scope}

\end{tikzpicture}
    }
    \caption{The vicinity of the stick-slip singularity in the physical $(x,y)$-plane with a polar-coordinate overlay $(r,\theta)$ centered at the singularity (left). The Newtonian flow is directed from right to left. The characteristic radial lines are the stick boundary at $\theta=0$, the core line at $\theta=\theta_c$, and the slip boundary at $\theta=\pi$; the two shades of purple mark the regions of inward ($\theta<\theta_c$) and outward ($\theta>\theta_c$) radial flow separated by the core line. On the right, the same domain is mapped to the $(\ln r,\theta)$-plane: the singularity corresponds to $\ln r\to-\infty$ on the left, the slip boundary is on top, each radial line becomes a horizontal line, and the streamlines are distorted by the logarithmic scaling of the radial axis.}
    \label{fig:stickslip_kartesian_to_polar_illu}
\end{figure*}
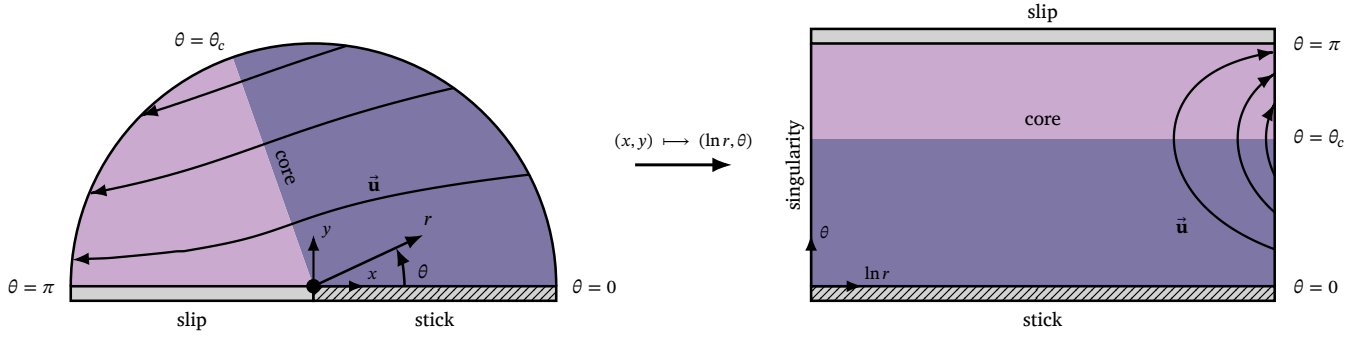

The Newtonian Stokes flow near the stick-slip singularity is described by the stream function $\tilde{\psi} = Cr^{3/2}\sin\frac{\theta}{2}\sin\theta$~\cite{richardson1970stick}, from which the polar velocity components follow as
\begin{align}
    u_r &= \frac{1}{r}\partial_\theta \tilde{\psi} = C r^{1/2} \sin \frac{\theta}{2} \left(2-3\sin^2 \frac{\theta}{2}\right) \label{eqn:ur}\\
    u_\theta &= -\partial_r \tilde{\psi}  = -\frac{3}{2}\, C r^{1/2} \sin \frac{\theta}{2} \sin \theta\, , \label{eqn:utheta}
\end{align}
and the polar-coordinate representation of the velocity
gradient as
\begin{gather}
    \widetilde{\nabla\mathbf{u}} =
    C r^{-1/2}
    \begin{pmatrix}[1.8]
        \frac{1}{2} \sin \frac{\theta}{2}
            \left(2-3\sin^2 \frac{\theta}{2}\right) &
        -\frac{3}{4} \sin \frac{\theta}{2} \sin \theta
            + \cos \frac{\theta}{2} \\
        -\frac{3}{4} \sin \frac{\theta}{2} \sin \theta &
        -\frac{1}{2} \sin \frac{\theta}{2}
            \left(2-3\sin^2 \frac{\theta}{2}\right)
    \end{pmatrix}.
\end{gather}
The radial velocity~\eqref{eqn:ur} vanishes at $\theta_c \coloneqq 2\arcsin\bigl(\sqrt{2/3}\bigr) \approx 1.91$, which defines the unique radial line to which all streamlines are orthogonal. This core line naturally bisects the domain: for $\theta < \theta_c$ the flow is directed toward the singularity, and for $\theta > \theta_c$ away from it. An illustration of the neighborhood of the stick-slip singularity, showing the Newtonian Stokes flow, the core line, and the transformation to polar coordinates, is given in Fig.~\ref{fig:stickslip_kartesian_to_polar_illu}.
In the given velocity field, the Giesekus fluid develops specific boundary layers at the stick and slip walls, and the upcoming asymptotic analysis is based on a separation of the stick-slip domain into three main regions: a stick boundary region with $\theta$ close to $0$, a core region at sufficient distance from both walls at a given radial distance, and a slip region with $\theta$ close to $\pi$.

With the velocity field now specified, the generic vorticity and strain tensors appearing in the polar log-conf Eq.~\eqref{eqn:polar_logconf} of Section~\ref{sec:log_conf_polar} take the explicit forms
\begin{align*}
    \tilde{\omega}(\mathbf{u}) =& \mathbf{O}^T \frac{\nabla\mathbf{u} - \nabla\mathbf{u}^T}{2} \mathbf{O} = \frac{\widetilde{\nabla\mathbf{u}} - \widetilde{\nabla\mathbf{u}}^T}{2} = \, \frac{C}{r^{1/2}} \begin{pmatrix} 0 & \omega_{r\theta} \\ -\omega_{r\theta} & 0 \end{pmatrix}\\
    &\qquad \omega_{r\theta} = \, \frac{1}{2} \cos \frac{\theta}{2}\\
    \tilde{\epsilon}(\mathbf{u}) =& \mathbf{O}^T \frac{\nabla\mathbf{u} + \nabla\mathbf{u}^T}{2} \mathbf{O} = \frac{\widetilde{\nabla\mathbf{u}} + \widetilde{\nabla\mathbf{u}}^T}{2} = \, \frac{C}{r^{1/2}} \begin{pmatrix} \epsilon_{rr} & \epsilon_{r\theta} \\ \epsilon_{r\theta} & -\epsilon_{rr} \end{pmatrix}\\
    &\qquad \epsilon_{rr} = \, \frac{1}{2} \sin \frac{\theta}{2}\left(2-3\sin^2 \frac{\theta}{2}\right)\\
    &\qquad \epsilon_{r\theta} = \, - \frac{3}{4} \sin \frac{\theta}{2} \sin \theta + \frac{1}{2} \cos \frac{\theta}{2}\, .
\end{align*}
Furthermore, we adopt $\ln r$ instead of $r$ as the radial coordinate, so that $\partial_{\ln r} = r\,\partial_r$, and the advective operator factorizes as
\begin{align*}
    u_r \partial_r + \frac{u_\theta}{r} \partial_\theta = \frac{C}{r^{1/2}}\,\hat{\partial}\,, \qquad
    \hat{\partial} \coloneqq \hat{u}_r\,\frac{\partial}{\partial \ln r}
        + \hat{u}_\theta\,\frac{\partial}{\partial \theta}\, ,
\end{align*}
with the velocity profiles
\begin{align*}
    \hat{u}_r &= \frac{u_r}{C\,r^{1/2}}
        = \sin\frac{\theta}{2}\left(2 - 3\sin^2\frac{\theta}{2}\right) \\
    \hat{u}_\theta &= \frac{u_\theta}{C\,r^{1/2}}
        = -\frac{3}{2}\,\sin\frac{\theta}{2}\sin\theta\, .
\end{align*}
We also introduce the two tensors
\begin{align*}
    \hat{\omega}(\mathbf{u}) &= \frac{r^{1/2}}{C} \, \tilde{\omega}(\mathbf{u}) = \begin{pmatrix} 0 & \omega_{r\theta} \\ -\omega_{r\theta} & 0 \end{pmatrix} \\
    \hat{\epsilon}(\mathbf{u}) &= \frac{r^{1/2}}{C} \, \tilde{\epsilon}(\mathbf{u}) = \begin{pmatrix} \epsilon_{rr} & \epsilon_{r\theta} \\ \epsilon_{r\theta} & -\epsilon_{rr} \end{pmatrix}\, ,
\end{align*}
so that by substituting these expressions into the steady-state polar log-conf equation~\eqref{eqn:polar_logconf} and using the linearity of $f(\opad \mathbf{\tilde{\Psi}})$, we get
\begin{gather}\label{eqn:polar_logconf_rescaled}
\begin{split}
     \hat{\partial}\mathbf{\tilde{\Psi}} 
        = [\mathbf{\tilde{\Psi}}, \hat{u}_\theta \mathbf{A} - \hat{\omega}(\mathbf{u})] + 2 f(\opad \mathbf{\tilde{\Psi}}) \hat{\epsilon}(\mathbf{u}) - \frac{r^{1/2}}{C} \, P(e^{\mathbf{\tilde{\Psi}}})e^{-\mathbf{\tilde{\Psi}}}\, ,
\end{split}
\end{gather}
such that all constants carrying a physical dimension are bundled into a single factor.

For notational brevity, we drop the tilde on the second-order tensors in what follows, with the understanding that all components refer to the polar basis. Combining this formulation with the representation~\eqref{eqn:logconf-paper2014} of $f(\opad \mathbf{\Psi})\, \hat{\epsilon}(\mathbf{u})$ then yields
\begin{align*}
    \hat{\partial} \mathbf{\Psi} =&\;
        2\begin{pmatrix}\Psi_{r\theta} & - \gamma(\mathbf{\Psi}) \\
            -\gamma(\mathbf{\Psi}) & - \Psi_{r\theta}\end{pmatrix} \\
        &\qquad\quad\cdot\left(\hat{u}_\theta + \omega_{r\theta} - \left[\gamma(\mathbf{\Psi})\,\epsilon_{r\theta} - \Psi_{r\theta}\,\epsilon_{rr}\right]h(D)\right) \\
        &\qquad\quad + 2\hat{\epsilon}(\mathbf{u}) - \frac{r^{1/2}}{C} \, P(e^{\mathbf{\Psi}})\, e^{-\mathbf{\Psi}}\, .
\end{align*}
For the further analysis, we want to disentangle this matrix-valued PDE into a system of scalar PDEs, which is the main reason for considering the previously introduced quantities $\optr \mathbf{\Psi}$, $\sqrt{D}$, and $\beta$. For $\optr \mathbf{\Psi}$ it is immediately evident that
\begin{align}\label{eqn:trPsi_eqn}
\begin{split}
    \hat{\partial} \optr \mathbf{\Psi} =& -2 \lambda^{-1} C^{-1} r^{1/2} \left(1-2\alpha + \alpha \cosh\sqrt{D} e^{\optr\Psi/2}
            \right.\\&\qquad\left.-(1-\alpha)\cosh\sqrt{D} e^{-\optr\Psi/2}\right)\, .
\end{split}
\end{align}

In analyzing this equation, we will use techniques that are prototypical for what follows. In particular, we will show that the behavior near the singularity, i.e.\ for $\ln r \ll 0$, is mostly determined by the viscometric behavior next to the stick boundary, since any streamline that approaches the singularity sufficiently closely spends a considerable time of flight next to the stick boundary.

More concretely, using $\hat{u}_r = \theta + \mathcal{O}(\theta^3)$ and $\hat{u}_\theta = -\frac{3}{4} \theta^2 + \mathcal{O}(\theta^4)$, as well as $C < 0$, we can write the aforementioned equation for $\ln \theta \ll 0$ as
\begin{align}\label{eqn:trPsi_eqn2}
\begin{split}
    &- \frac{\partial \optr \mathbf{\Psi}}{\partial\ln r} + \frac{3}{4} \frac{\partial \optr \mathbf{\Psi}}{\partial\ln \theta}\\ =& -2 \lambda^{-1} \seminorm{C}^{-1} e^{\frac{1}{2}\ln r - \ln \theta} \Bigg[1-2\alpha + 2 \sqrt{\alpha(1-\alpha)} \cosh\sqrt{D}\\&\qquad\qquad\qquad\qquad\quad\quad\times \sinh\left(\frac{\optr\mathbf{\Psi}}{2} - \frac{1}{2} \ln \left(\frac{1-\alpha}{\alpha}\right)\right)\Bigg]\, .
\end{split}
\end{align}
Now, given a sufficient time of flight with $\frac{1}{2}\ln r - \ln\theta > 0$, $\optr\mathbf{\Psi}$ approaches the stable saturation limit
\begin{align}
    \optr\mathbf{\Psi} = \ln\left(\frac{1-\alpha}{\alpha}\right) + 2\arsinh\left(\frac{2\alpha-1}{2\sqrt{\alpha(1-\alpha)}\cosh\sqrt{D}}\right)\, ,
\end{align}
which is clearly bounded. Considering that some part of $\mathbf{\Psi}$ has to diverge to counter the singularity of $\epsilon(\mathbf{u})$ in~\eqref{eqn:logconf} or of $\tilde{\epsilon}(\mathbf{u})$ in~\eqref{eqn:polar_logconf}, respectively, the only degree of freedom left is $\sqrt{D}$, which, as we will also see a posteriori, satisfies $\sqrt{D} \to\infty$ for $\ln r \to -\infty$.

Therefore, it is reasonable to assume that $\optr\mathbf{\Psi}$ is asymptotically given by
\begin{align}\label{eqn:soltrpsi}
    \optr\mathbf{\Psi} = \ln\left(\frac{1-\alpha}{\alpha}\right)\, .
\end{align}

\begin{rem}
It is worth noting that the constant trace in the asymptotic solution is a rather unique feature of our derivation. The existing literature has, at least in part, noticed an upper bound~\cite[Eq.~(9)]{renardy1997high2} and used it to justify a low-rank approximation of $\mathbf{C}$. The latter would of course blow up $\mathbf{\Psi}$ and is thus unsuitable for our analysis. However, there do not seem to be any fundamental obstructions to preserving the trace of $\mathbf{\Psi}$, or the determinant of $\mathbf{C}$, in non-log-conf formulations; this may simply have gone unnoticed. For instance, the solution in~\cite[Eqs.~(3.41)--(3.45)]{evans2015stick} can easily be shown to fulfill this property.
\end{rem}

To derive similar equations for $\sqrt{D}$ and $\tilde{\beta}$, we first derive, as an intermediate step, the equations for $\Psi_{r\theta}$ and $\gamma(\mathbf{\Psi})$:
\begin{align*}
    \hat{\partial} \Psi_{r\theta} =& - 2 \gamma(\mathbf{\Psi})\left(\hat{u}_\theta+\omega_{r\theta} - \left[\gamma(\mathbf{\Psi})\epsilon_{r\theta}
        -\Psi_{r\theta}\epsilon_{rr}\right]h(D)\right)\\
        & + 2 \epsilon_{r\theta} - (1-\alpha) \frac{r^{1/2}}{C\lambda} \Psi_{r\theta} \frac{\sinh\sqrt{D}}{\sqrt{D}}
                e^{-\optr\mathbf{\Psi}/2}\\
        & - \alpha \frac{r^{1/2}}{C\lambda} \Psi_{r\theta} \frac{\sinh\sqrt{D}}{\sqrt{D}} e^{\optr\mathbf{\Psi}/2} \\
    \hat{\partial} \gamma(\mathbf{\Psi}) =& 2 \Psi_{r\theta} \left(\hat{u}_\theta+\omega_{r\theta} - \left[\gamma(\mathbf{\Psi})\epsilon_{r\theta}
        -\Psi_{r\theta}\epsilon_{rr}\right]h(D)\right)\\
        & + 2 \epsilon_{rr} - (1-\alpha) \frac{r^{1/2}}{C\lambda} \gamma(\mathbf{\Psi}) \frac{\sinh\sqrt{D}}{\sqrt{D}}
                e^{-\optr\mathbf{\Psi}/2}\\
        & - \alpha \frac{r^{1/2}}{C\lambda} \gamma(\mathbf{\Psi}) \frac{\sinh\sqrt{D}}{\sqrt{D}} e^{\optr\mathbf{\Psi}/2} \, .
\end{align*}
It then follows for $\tilde{\beta}$ that
\begin{align*}
    \hat{\partial} \tilde{\beta} =& \frac{-\Psi_{r\theta}}{D} \hat{\partial} \gamma(\mathbf{\Psi})
                + \frac{\gamma(\mathbf{\Psi})}{D} \hat{\partial}\Psi_{r\theta}\\
        =& -2\left(\hat{u}_\theta+\omega_{r\theta} - \left[\gamma(\mathbf{\Psi})\epsilon_{r\theta}
        -\Psi_{r\theta}\epsilon_{rr}\right]\left(h(D)+\frac{1}{D}\right)\right)\\
        =& -2\left(\hat{u}_\theta+\omega_{r\theta} - \coth{\sqrt{D}} \left[\cos\tilde{\beta}\, \epsilon_{r\theta}
        -\sin\tilde{\beta}\, \epsilon_{rr}\right]\right)\, ,
\end{align*}
where the last step uses the definition of $h$ as well as $\sqrt{D}\cos\tilde{\beta} = \gamma(\mathbf{\Psi})$ and $\sqrt{D}\sin\tilde{\beta} = \Psi_{r\theta}$.

For the Cartesian $\beta = \tilde{\beta} + 2\theta$, we have the equation
\begin{align}\label{eqn:betatildeeqn}
    \hat{\partial}\beta =& -2\left(\omega_{r\theta} - \coth{\sqrt{D}} \left[\cos(\beta-2\theta)\, \epsilon_{r\theta}
        -\sin(\beta-2\theta)\, \epsilon_{rr}\right]\right)\, .
\end{align}

Similarly, for $\hat{\partial}\sqrt{D}$ we obtain
\begin{align*}
    \hat{\partial} \sqrt{D} =& D^{-1/2} \left(\gamma(\mathbf{\Psi})\hat{\partial}\gamma(\mathbf{\Psi}) + \Psi_{r\theta}\hat{\partial}\Psi_{r\theta}\right)\\
        =& 2(\cos\tilde{\beta} \epsilon_{rr} + \sin\tilde{\beta} \epsilon_{r\theta})\\
        & - (1-\alpha) \frac{r^{1/2}}{C\lambda} e^{-\optr\mathbf{\Psi}/2} \sinh\sqrt{D}\\
        & - \alpha \frac{r^{1/2}}{C\lambda} e^{\optr\mathbf{\Psi}/2} \sinh\sqrt{D}\, .
\end{align*}

Since $\optr\mathbf{\Psi} \to \ln\left(\frac{1-\alpha}{\alpha}\right)$ as we move toward the singularity, we can simplify the $\sqrt{D}$ equation further
\begin{align}
\begin{split}
    \hat{\partial} \sqrt{D} =& 2(\cos\tilde{\beta}\, \epsilon_{rr} + \sin\tilde{\beta}\, \epsilon_{r\theta})\\
        & - \sqrt{\alpha(1-\alpha)}\, \frac{2r^{1/2}}{C\lambda} \sinh\sqrt{D}\, .
\end{split}
\end{align}

Introducing
\begin{align}\label{eqn:gamma_0}
    \gamma_0 = \frac{1}{2} \ln \frac{\sqrt{\alpha(1-\alpha)}}{2 |C|\lambda}\, ,
\end{align}
and noting that $C < 0$, we can further write
\begin{align}\label{eqn:asympsqrtDeqn}
\begin{split}
    \hat{\partial} \sqrt{D} =& 2(\cos\tilde{\beta}\, \epsilon_{rr} + \sin\tilde{\beta}\, \epsilon_{r\theta})\\
        & + 4 e^{\frac{1}{2} \ln r + 2 \gamma_0} \sinh\sqrt{D}\, .
\end{split}
\end{align}

We have thus essentially arrived at the two equations~\eqref{eqn:betatildeeqn} and~\eqref{eqn:asympsqrtDeqn}, for which we seek an asymptotic solution in the following sections.

\begin{rem}[Weissenberg number]
For a given $r$, one can argue that $\seminorm{C}/r^{1/2}$ is a strain rate and thus $\text{Wi} = \frac{\lambda \seminorm{C}}{r^{1/2}}$ constitutes the Weissenberg number, which clearly grows without bound as one approaches the singularity. This also strongly reflects the fact that in the core region, which we will study in Section~\ref{sec:core}, the dynamics are asymptotically governed by those of a UCM fluid.

The constant $\gamma_0$, which strictly speaking carries the logarithm of a physical dimension, takes over in our setting the role of characterizing the singular solution in terms of the problem-dependent parameters alone.
\end{rem}

\section{Stick boundary layer and transition zone}

In the following, we take a detailed look at the extremely sharp boundary layer at the stick boundary, which is instrumental in determining the slope of $\sqrt{D}$ in the core region.

We perform a small-angle approximation in $\theta$:
\begin{align}
    \omega_{r\theta} =& \frac{1}{2} - \frac{1}{16} \theta^2 + \mathcal{O}(\theta^4)\\
    \epsilon_{rr} =& \frac{\theta}{2} + \mathcal{O}(\theta^3)\\
    \epsilon_{r\theta} =& \frac{1}{2} - \frac{7}{16}\theta^2 + \mathcal{O}(\theta^3)\\
    \hat{u}_r =& \theta + \mathcal{O}(\theta^3)\\
    \hat{u}_\theta =& -\frac{3}{4} \theta^2 + \mathcal{O}(\theta^4)\, .
\end{align}

This yields the following equation for $\beta$
\begin{align*}
    -\frac{1}{2} \hat{\partial} \beta =& \omega_{r\theta} - \coth\sqrt{D} \left[\cos(\beta-2\theta)\epsilon_{r\theta} - \sin(\beta-2\theta)\epsilon_{rr}\right]\\
        =& \frac{1}{2} - \frac{1}{16}\theta^2 - \coth\sqrt{D}\left[\left((1-2\theta^2)\cos\beta+2\theta\sin\beta\right)\right.\\
        &\qquad\left.\left(\frac{1}{2}-\frac{7}{16}\theta^2\right) - (\sin\beta-2\theta\cos\beta)\frac{\theta}{2}\right]+\mathcal{O}(\theta^3)\\
        =& \frac{1}{2} - \frac{1}{16}\theta^2 - \coth\sqrt{D}\left[\frac{1}{2}\cos\beta +\frac{1}{2}\theta\sin\beta\right.\\&\qquad\left.- \frac{7}{16}\theta^2\cos\beta\right]+\mathcal{O}(\theta^3)\, .
\end{align*}
Applying the Weierstraß substitution $\tilde{u} = \tan\frac{\beta}{2}$ with $\partial_i \beta = \frac{2\partial_i \tilde{u}}{1+\tilde{u}^2}$, $\cos\beta = \frac{1-\tilde{u}^2}{1+\tilde{u}^2}$, and $\sin\beta = \frac{2\tilde{u}}{1+\tilde{u}^2}$ yields
\begin{align*}
    -\hat{\partial} \tilde{u} =& \left(\frac{1}{2} - \frac{1}{16}\theta^2\right) (1+\tilde{u}^2) - \coth\sqrt{D}\left[\frac{1}{2}(1-\tilde{u}^2) \right.\\
        &\qquad\left. + \theta\tilde{u} -\frac{7}{16}\theta^2 (1-\tilde{u}^2) \right] +\mathcal{O}(\theta^3)\, .
\end{align*}

Note that the leading terms in the derivative start at order $\theta$; hence, for small $\theta$, the zeroth-order term, which depends heavily on $\sqrt{D}$, dominates the solution. For this reason, we are only allowed to use the $\sqrt{D} \gg 1$ approximation $\coth\sqrt{D} \approx 1$ in the higher-order terms. This then gives
\begin{align*}
    -\hat{\partial} \tilde{u} =& \tilde{u}^2 - e^{-2\sqrt{D}} - \theta\tilde{u} + \frac{3}{8}\theta^2 + \mathcal{O}(\theta^3,\theta^2\tilde{u}^2)\, .
\end{align*}
Reformulating in terms of $\ln\theta$, we obtain
\begin{align*}
    -e^{\ln\theta} \left(\frac{\partial\tilde{u}}{\partial \ln r} - \frac{3}{4} \frac{\partial\tilde{u}}{\partial \ln\theta}\right) =& \tilde{u}^2 - e^{-2\sqrt{D}} - e^{\ln\theta} \tilde{u} + \frac{3}{8}e^{2\ln\theta}\, ,
\end{align*}
which, upon introducing $\tilde{\tilde{u}} = e^{-\ln\theta} \tilde{u}$, yields
\begin{align*}
    - \frac{\partial\tilde{\tilde{u}}}{\partial\ln r} + \frac{3}{4} \frac{\partial\tilde{\tilde{u}}}{\partial\ln\theta}
        =& - e^{- \ln\theta}\left(\frac{\partial\tilde{u}}{\partial \ln r} - \frac{3}{4} \frac{\partial\tilde{u}}{\partial \ln\theta}\right) - \frac{3}{4} e^{- \ln\theta} \tilde{u}\\
        =& \tilde{\tilde{u}}^2 - e^{-2\sqrt{D}-2\ln\theta} - \frac{7}{4} \tilde{\tilde{u}} + \frac{3}{8}\, .
\end{align*}
In the following, we apply similar steps to the $\sqrt{D}$-equation~\eqref{eqn:asympsqrtDeqn}. Starting with the small-angle approximation
and assuming $\sqrt{D}$ to be already large enough for $\sinh\sqrt{D} \approx e^{\sqrt{D}}/2$, this yields
\begin{align*}
    \hat{\partial} \sqrt{D} =& 2 \left[((1-2\theta^2)\cos\beta+2\theta\sin\beta)\frac{\theta}{2}
                \right.\\&\qquad\left.+ (\sin\beta-2\theta\cos\beta)\left(\frac{1}{2}-\frac{7}{16}\theta^2\right)\right]
            \\& +2 e^{\frac{1}{2}\ln r + \gamma_0} e^{\sqrt{D}+\gamma_0} + \mathcal{O}(\theta^3)\, ,
\end{align*}
which, after applying the same Weierstraß substitution as above, leads to
\begin{align*}
    \hat{\partial} \sqrt{D} =& \frac{2}{1+\tilde{u}^2} \left[((1-2\theta^2)(1-\tilde{u}^2)+4\theta\tilde{u})\frac{\theta}{2}
                \right.\\&\qquad\left.+ (2\tilde{u}-2\theta(1-\tilde{u}^2))\left(\frac{1}{2}-\frac{7}{16}\theta^2\right)\right]
            \\& + 2e^{\frac{1}{2}\ln r + \gamma_0} e^{\sqrt{D}+\gamma_0} + \mathcal{O}(\theta^3)\\
        =& 2 \left[-\frac{\theta}{2} + \tilde{u}\right]
                + 2e^{\frac{1}{2}\ln r + \gamma_0} e^{\sqrt{D}+\gamma_0}\\
            &  + \mathcal{O}(\theta^3,\theta^2 \tilde{u}, \theta\tilde{u}^2)\, .
\end{align*}
In terms of $\ln\theta$, this yields
\begin{align*}
    - \frac{\partial\sqrt{D}}{\partial\ln r} + \frac{3}{4} \frac{\partial\sqrt{D}}{\partial\ln\theta}
        =& 1 - 2\tilde{\tilde{u}} - 2e^{\frac{1}{4}\ln r -\ln\theta + \gamma_0} e^{\sqrt{D}+\frac{1}{4}\ln r + \gamma_0}\, .
\end{align*}
Introducing
\begin{align}\label{eqn:sqrtD_sol_stick}
    g = \sqrt{D} + \frac{1}{4}\ln r + \gamma_0\, ,
\end{align}
we obtain
\begin{align*}
    - \frac{\partial g}{\partial\ln r} + \frac{3}{4} \frac{\partial g}{\partial\ln\theta}
        =& \frac{3}{4} - 2\tilde{\tilde{u}} - 2 e^{\frac{1}{4}\ln r -\ln\theta +\gamma_0} e^{g}\\
    - \frac{\partial\tilde{\tilde{u}}}{\partial\ln r} + \frac{3}{4} \frac{\partial\tilde{\tilde{u}}}{\partial\ln\theta}
        =& \tilde{\tilde{u}}^2 - e^{-2g+2\left(\frac{1}{4}\ln r -\ln\theta +\gamma_0\right)} - \frac{7}{4} \tilde{\tilde{u}} + \frac{3}{8}\, .
\end{align*}
Note that the right-hand sides depend only on
\begin{align}\label{eqn:def_x}
    x\coloneqq -\frac{1}{4}\ln r +\ln\theta -\gamma_0\, ,
\end{align}
which encodes the stick transition. Hence, assuming that the initial or boundary conditions of our problem are sufficiently homogeneous ($g$ starting from the same value regardless of the concrete starting point $(r,\theta)$ of the integration), the solution will depend on $x$ only. That this is indeed the case will be justified in the following.

We make the following ansatz
\begin{align}
    g(\ln r, \ln\theta) =& \tilde{g} (-\frac{1}{4}\ln r +\ln\theta -\gamma_0)\\
    \tilde{\tilde{u}}(\ln r, \ln\theta) =& \tilde{\tilde{\tilde{u}}} (-\frac{1}{4}\ln r +\ln\theta -\gamma_0)\,.
\end{align}
Therefore, we arrive at the coupled ODE system
\begin{align}\label{eqn:ode_gtilde}
\begin{split}
    \frac{d\tilde{g}}{dx} =& \frac{3}{4} - 2\tilde{\tilde{\tilde{u}}} - 2 e^{-x} e^{\tilde{g}}\\
    \frac{d\tilde{\tilde{\tilde{u}}}}{dx} =& - e^{-2\tilde{g}-2x} + \left(\tilde{\tilde{\tilde{u}}} - \frac{1}{4}\right)\left(\tilde{\tilde{\tilde{u}}} - \frac{3}{2}\right) \, .
\end{split}
\end{align}
Introducing $l$ such that $\tilde{\tilde{\tilde{u}}} = \frac{3}{8}-e^{-x}l$, we can also reformulate the system as
\begin{align}
\begin{split}
    \frac{d\tilde{g}}{dx} =& 2 e^{-x}\left(l - e^{\tilde{g}}\right)\\
    \frac{dl}{dx} =& e^{-x}\left(e^{-2\tilde{g}}-l^2\right)+\frac{9}{64} e^x \, .
\end{split}
\end{align}
For $x\ll 0$, it is immediately evident that, whatever the initial condition, the system is driven extremely quickly to the states $l^2 = e^{-2\tilde{g}}$ and $l = e^{\tilde{g}}$, whose only common solution is $\tilde{g} = 0$ and $l=1$, thus justifying our above assumption on the initial conditions.

For $x\gg 0$, one can easily verify by substitution that the following ansatz yields an asymptotic and stable solution
\begin{align}\label{eqn:approxgtilde}
    \tilde{g}(x) \approx& \frac{x}{4}+a_1\\
    \tilde{\tilde{\tilde{u}}}(x) \approx& \frac{1}{4}\, ,
\end{align}
with some integration constant $a_1$.

\begin{rem}\label{rem:stick_second_solution}
In principle, the factorization in~\eqref{eqn:ode_gtilde} suggests that $\tilde{\tilde{\tilde{u}}} = \frac{3}{2}$ might also be a solution. Apart from the fact that this is the unstable choice of the two roots, it would initially lead to $\frac{d\tilde{g}}{dx} = -\frac{9}{4} < -1$, such that $-e^{-2\tilde{g}-2x}$ becomes significant again at some point, thus causing $\tilde{\tilde{\tilde{u}}}$ to decrease and driving the whole system back to the $\tilde{\tilde{\tilde{u}}}=\frac{1}{4}$ solution.
\end{rem}

\begin{figure}[h!]
    \centering
     \includegraphics[width=\linewidth]{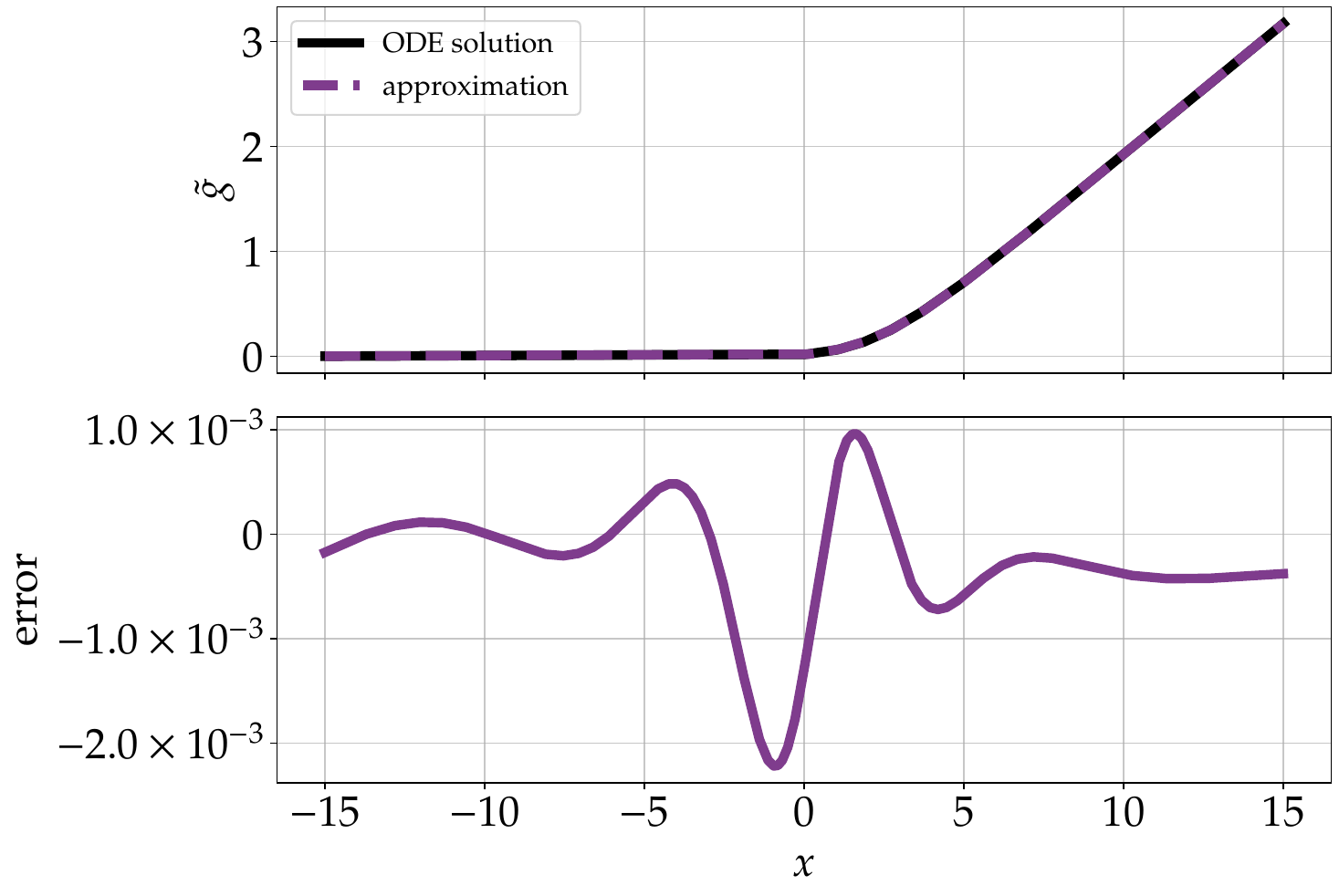}
    \caption{The function $\tilde{g}$ and its approximation error.}
    \label{fig:gfunc_final}
\end{figure}

Returning to the original variables $\sqrt{D}$ and $\beta$, this yields for $-\frac{1}{4}\ln r + \ln\theta -\gamma_0 \ll 0$ the solutions
\begin{align}\label{eqn:sqrtD_before_stick_trans}
    \sqrt{D} \approx& - \frac{1}{4} \ln r - \gamma_0 \\
    \beta \approx& \frac{3}{4}\theta - 2 e^{\frac{1}{4}\ln r + \gamma_0}\,.\label{eqn:beta_before_stick_trans}
\end{align}
Similarly, for $-\frac{1}{4}\ln r + \ln\theta -\gamma_0 \gg 0$,
\begin{align}\label{eqn:sqrtD_after_stick_trans}
    \sqrt{D} \approx& - \frac{5}{16} \ln r + \frac{1}{4}\ln \theta - \frac{5}{4} \gamma_0 + a_1 \\
    \beta \approx& \frac{1}{2}\theta\,.\label{eqn:beta_after_stick_trans}
\end{align}
However, connecting these two limits requires the full solution of the ODE
system~\eqref{eqn:ode_gtilde}, which we obtain numerically with an in-house implicit BDF2 solver implemented in PyTorch~\cite{paszke2019pytorch}. In order to obtain a closed-form representation, the numerical solution is then fitted by 
\begin{align}\label{eqn:pade_gtilde}
    \tilde{g}(x) \approx 2\ln\left(\frac{p(e^{-x/8})}{q(e^{-x/8})}\right)\, ,
\end{align}
with the argument of the logarithm being a rational function of $t = e^{-x/8}$, whose numerator $p(t)$ and denominator $q(t)$ are polynomials of degree eight in the monomial basis with coefficients $p_i$ and $q_i$, respectively. The coefficients are constrained to be non-negative by optimizing over the logarithms of the coefficients, $\{\ln p_i,\ln q_i\}$, with $p_8 = q_8 = 1$ and $q_0 = 0$ held fixed; this ensures $p,q>0$ for $t > 0$ and hence a pole-free, well-defined approximation on the whole real axis. The former constraint ensures $\tilde{g}\to 0$ for $x\to-\infty$, and the latter is imposed so that the linear growth $\tilde{g}\sim x/4$ of Eq.~\eqref{eqn:approxgtilde} is reproduced exactly. The coefficients are determined by minimizing a weighted squared deviation from the numerical solution in PyTorch. The coefficients of both polynomials are given in Tab.~\ref{tab:pade_coeffs}, and $\tilde{g}$ is shown in Fig.~\ref{fig:gfunc_final}. The constant $a_1$ is then computed via
\begin{align}\label{eqn:a1_constant}
a_1 = 2\ln\!\left(\frac{p_0}{q_1}\right) \approx -0.578\, ,
\end{align}
where $p_0$ is the constant coefficient of $p$ and $q_1$ the linear coefficient of $q$. This value, obtained from the rational-function fit, is also consistent with the one obtained directly from the ODE solution.
\begin{table}[h!]
    \centering
    \begin{tabular*}{\columnwidth}{@{\extracolsep{\fill}}c*{2}{S[table-format=1.10]}@{}}
    \toprule
    $k$ & {$p_k$} & {$q_k$} \\
    \midrule
    0 & 1.7427474698 & 0 \\
    1 & 0.2894369803 & 2.3269126162 \\
    2 & 0.1548055180 & 0.3815283448 \\
    3 & 0.2016368924 & 0.2291793273 \\
    4 & 1.7727219302 & 0.3516214954 \\
    5 & 0.8087940901 & 1.8432954642 \\
    6 & 1.2331874078 & 0.9735228514 \\
    7 & 2.7724447271 & 2.7938282114 \\
    8 & 1 & 1 \\
    \bottomrule
    \end{tabular*}
    \caption{Coefficients of the rational representation~\eqref{eqn:pade_gtilde} in the monomial basis, $p(t) = \sum_k p_k t^k$ and $q(t) = \sum_k q_k t^k$ with $t = e^{-x/8}$, rounded to ten decimal places.}
    \label{tab:pade_coeffs}
\end{table}

The solution for $\tilde{\tilde{\tilde{u}}}$ can then be recovered by rearranging~\eqref{eqn:ode_gtilde}:
\begin{align}\label{eqn:approx_tildeu_stick}
    \tilde{\tilde{\tilde{u}}}(x) = \frac{3}{8}-e^{-x}e^{\tilde{g}} - \frac{1}{2} \frac{d\tilde{g}}{dx}\, .
\end{align}

\section{Core region}\label{sec:core}
The core region is special in that the $P(e^{\mathbf{\Psi}})$ term is suppressed and the flow is essentially governed by the same equations as a UCM fluid.

Inspired by the previous results in the stick region, we introduce the field
\begin{align}\label{eqn:first_def_o}
\begin{split}
    o(\ln r, \theta) =& \sqrt{D} + \frac{1}{4}\ln r + \gamma_0\\&\quad -\tilde{g}\left(-\frac{1}{4}\ln r + \ln(2\,\sin(\theta/2)) - \gamma_0\right)\\&\quad
                        + \frac{7}{8} \ln (\cos(\theta/2))\,,
\end{split}
\end{align}
which for small $\theta$ can be approximated by
\begin{align}
    o(\ln r, \theta) \approx& \sqrt{D} + \frac{1}{4}\ln r + \gamma_0 -\tilde{g}\left(-\frac{1}{4}\ln r + \ln(\theta) - \gamma_0\right)\,,
\end{align}
such that $o(\ln r, 0) = 0$ is fulfilled; in particular, the logarithmic singularity at $\theta=0$ has been absorbed into the definition of $o$.

The inclusion of the $\cos(\theta/2)$ term is motivated by a similar argument, which, however, anticipates the next section: we want to absorb a logarithmic singularity at the slip boundary that has a prefactor of $7/8$. The particular choice of $\cos(\theta/2)$ is motivated by the requirement that this term must not interfere with the boundary conditions at the stick boundary.

In the core region, we can further approximate $\tilde{g}$ as given above in Eq.~\eqref{eqn:approxgtilde}. Hence, we have
\begin{align}\label{eqn:approx_def_o_core}
\begin{split}
    o(\ln r, \theta) =& \sqrt{D} + \frac{5}{16}\ln r + \frac{5}{4} \gamma_0 -\frac{1}{4} \ln(2\,\sin(\theta/2))\\&\quad - a_1
                        + \frac{7}{8} \ln (\cos(\theta/2))\, .
\end{split}
\end{align}
Using the defining equation for $\sqrt{D}$, i.e.\ Eq.~\eqref{eqn:asympsqrtDeqn}, then yields
\begin{align}\label{eqn:core_o_eqn}
\begin{split}
    \hat{u}_\theta\frac{d\, o}{d\, \theta} =& \frac{5}{16} \hat{u}_r - \frac{1}{8} \frac{\cos(\theta/2)}{\sin(\theta/2)} \hat{u}_\theta 
                - \frac{7}{16} \frac{\sin(\theta/2)}{\cos(\theta/2)}\hat{u}_\theta\\&\quad
            + 2(\cos(\beta-2\theta)\, \epsilon_{rr} + \sin(\beta-2\theta)\, \epsilon_{r\theta})\, .
\end{split}
\end{align}
Note that we have omitted the $r^{1/2} \sinh\sqrt{D}$ term, since it is exponentially suppressed in $\ln r$ until we come sufficiently close to the slip boundary, where the $-\frac{7}{8} \ln (\cos(\theta/2))$ singularity becomes dominant. This will be discussed in the following section. Given the boundary conditions at the stick boundary, we have a $\ln r$-independent solution, which allows us to drop the $\partial o/\partial\ln r$ term and to consider $o(\theta)$ as a function of $\theta$ only.

We still need an ODE for $\beta$ in the core region, which, using $\coth \sqrt{D} \approx 1$, is just the following variant of~\eqref{eqn:betatildeeqn}:
\begin{align}\label{eqn:core_betatilde_eqn}
    \hat{u}_\theta\frac{d\beta}{d\theta} =& -2\left(\omega_{r\theta} - \cos(\beta-2\theta)\, \epsilon_{r\theta}
        +\sin(\beta-2\theta)\, \epsilon_{rr}\right)\, .
\end{align}
On the one hand, \eqref{eqn:beta_after_stick_trans} clearly motivates the initial condition $\beta = 0$ at $\theta=0$; on the other hand, as one can see, this is also the only sensible choice: $\hat{u}_\theta$ vanishes at $\theta=0$, such that the equation becomes purely algebraic in the vicinity of $\theta=0$, and using the definitions of $\omega_{r\theta}$ etc., one can easily check that $\beta = 0$ (modulo multiples of $2\pi$) is the only solution when equating the right-hand side to zero. Note, however, that $\hat{u}_\theta$ has a double zero at $\theta=0$, such that fixing $\beta(0)=0$ is not enough, and the slope needs to be determined as well, as we will see shortly.

In the following, we show that Eqs.~\eqref{eqn:core_o_eqn} and~\eqref{eqn:core_betatilde_eqn} can be solved in closed form.
The starting point is once again the Weierstraß substitution $\tilde{u} = \tan\frac{\beta}{2}$. Furthermore, we divide the equation by $\sin\frac{\theta}{2}$ and also apply the Weierstraß substitution $t = \tan\frac{\theta}{2}$, which yields (with the help of a computer algebra system such as SymPy~\cite{sympy}) the following algebraic differential equation for $\tilde{u}$:
\begin{align}
    \frac{d\tilde{u}}{dt} =& \frac{-t^4 + 3t^2+\tilde{u}^2\left(3t^4 +t^2+2\right)+ \tilde{u}\left(-2t^5+2t^3-4t\right)}{3t^2\left(t^4+2t^2+1\right)}\,.
\end{align}
Note that the latter is a Riccati equation with rational coefficients, which can be solved with SymPy's implementation of Kovacic's algorithm (cf.~\cite[Algorithm~11]{vo2016rational} and~\cite{KOVACIC19863}). This gives the two solutions
\begin{align}\label{eqn:first_sol_utilde_core}
    \tilde{u}_1(t) =& \frac{t}{t^2+2}\\
    \tilde{u}_2(t) =& \frac{5t^3 + 3t}{3t^2 + 1}\, .\label{eqn:second_sol_utilde}
\end{align}

In order to match the slope $\tilde{u}'(t) = \tfrac{1}{2}$ from the stick regime, we select $\tilde{u}(t) = \tilde{u}_1(t)$, which after back-substitution yields
\begin{align}\label{eqn:sol_utilde_core}
    \tilde{u}(\theta) =& \frac{\sin \theta}{3+\cos\theta}
\end{align}
and
\begin{align}\label{eqn:sol_beta_core}
    \beta(\theta) =& 2\arctan\left(\frac{\sin \theta}{3+\cos\theta}\right)\,.
\end{align}

The other solution, $\tilde{u}_2(t)$, exactly matches the second solution from Remark~\ref{rem:stick_second_solution}, which was found there to be unstable in our current viscometric setup.

Now, considering the equation for $o(\theta)$, we apply the same machinery: dividing by $\sin\frac{\theta}{2}$, applying the same substitution of $\theta$ by $t$, and using the solution for $\tilde{u}$, we obtain for $o(t)$ the following algebraic differential equation
\begin{align}
    \frac{do}{dt} =& \frac{2t}{t^2+4} - \frac{2t}{t^2+1}\, .
\end{align}
In this partial-fraction form, the solution with $o(0)=0$ is readily obtained as
\begin{align}
    o(t) =& -\ln(t^2+1)+\ln(t^2+4)-\ln 4\, .
\end{align}
Back-substitution then yields
\begin{align}
    o(\theta) =& \ln\left(\frac{5+3\cos\theta}{8}\right)\, . \label{eqn:sol_o_core}
\end{align}

\begin{figure}[h!]
    \centering
     \includegraphics[width=\linewidth]{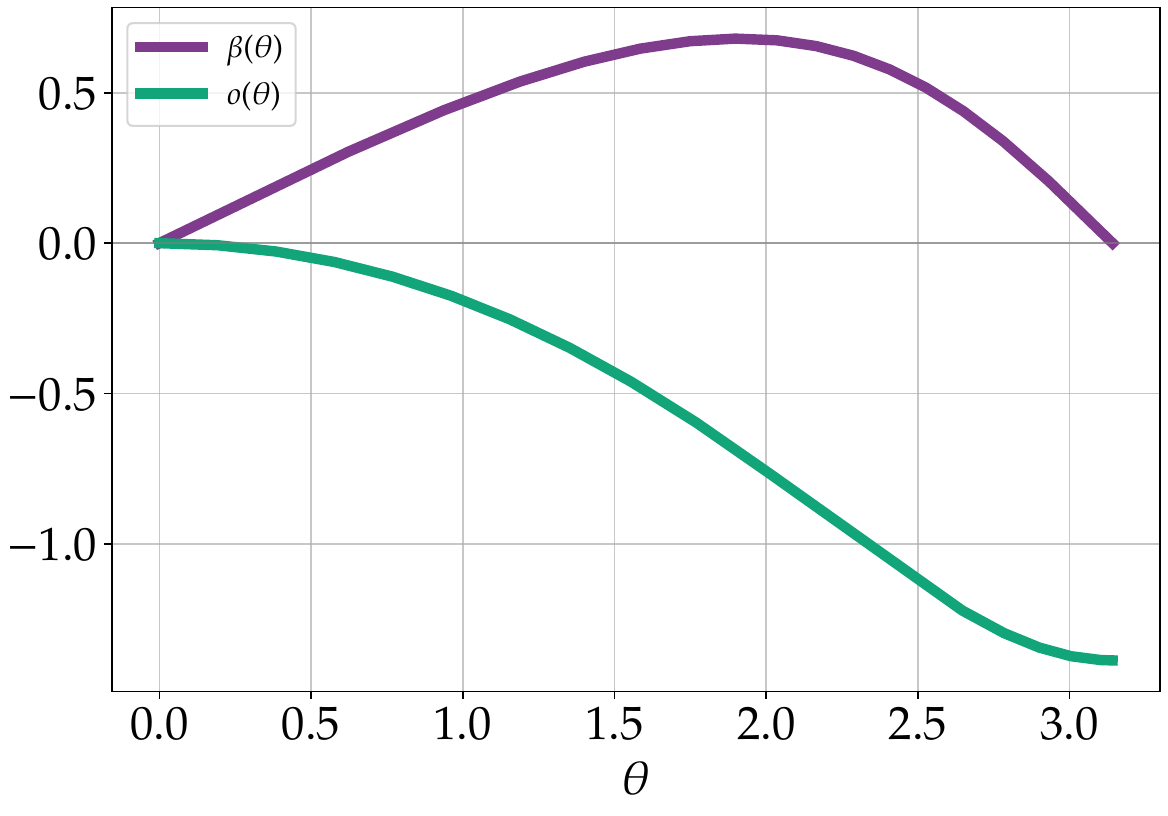}
    \caption{Analytical solutions $\beta$ and $o$ from~\eqref{eqn:sol_beta_core} and~\eqref{eqn:sol_o_core} for the core region as functions of $\theta$.}
    \label{fig:beta_o_plot}
\end{figure}

The solutions~\eqref{eqn:sol_beta_core} and~\eqref{eqn:sol_o_core} are shown in Fig.~\ref{fig:beta_o_plot}.

Thus, in combination with~\eqref{eqn:approx_def_o_core}, we have within the core region the following solution for $\sqrt{D}$
\begin{align}\label{eqn:sol_sqrtD_core}
\begin{split}
    \sqrt{D} =&  - \frac{5}{16}\ln r - \frac{5}{4} \gamma_0 + \frac{1}{4} \ln(2\,\sin(\theta/2))\\&\quad + a_1
                        - \frac{7}{8} \ln (\cos(\theta/2)) + o(\theta)\, .
\end{split}
\end{align}

\begin{rem}[Universality of the core solution and its relation to natural stress variables]
It is quite noteworthy that, for much of the analysis in this section, the specific assumption of the Giesekus model was not really necessary beyond providing initial values. In fact, for the most part we could equally have considered an Upper-Convected Maxwell~(UCM) fluid.

To stress this fact, and to make the connection to the extensive literature, we want to highlight the following: we are of course not limited to transforming $\mathbf{\Psi}$ to polar coordinates; based on~\eqref{eqn:rot_identity_beta}, we could choose any rotated coordinate system. One particularly important choice is the transformation to (scaled) natural stress variables~\cite{Renardy199491}, which in our notation corresponds to a rotation that maps $\mathbf{u}$ to $(\seminorm{\mathbf{u}},0)^T$. In this coordinate frame, the conformation tensor $\mathbf{C}$ is then given by
\begin{align}
    \begin{pmatrix} C_{uu} & C_{uw}\\ C_{uw} & C_{ww} \end{pmatrix}
        =& \begin{pmatrix} \lambda \seminorm{\mathbf{u}}^2 & \mu \\ \mu & \nu / \seminorm{\mathbf{u}}^2\end{pmatrix}\, ,
\end{align}
where $\lambda,\mu,\nu$ are the so-called natural stress variables. In~\cite{evans2019numerical,afonso2025natural}, $\hat{\lambda} = \lambda \seminorm{\mathbf{u}}^2$, $\hat{\mu} = \mu$, and $\hat{\nu} = \nu / \seminorm{\mathbf{u}}^2$ are referred to as scaled natural stress variables.

Note that the velocity components in the Cartesian basis, though parametrized by polar coordinates, can, owing to $\cos\theta = 1-2\sin^2\tfrac{\theta}{2}$, be written as
\begin{align*}
    u_x =& u_r \cos\theta - u_\theta \sin\theta\\
        =& \frac{1}{2} C r^{1/2} \sin \frac{\theta}{2} \left(3+\cos\theta\right)\\
    u_y =& u_r \sin\theta + u_\theta \cos\theta\\
        =& \frac{1}{2} C r^{1/2} \sin \frac{\theta}{2} \sin \theta\, .
\end{align*}
Hence, denoting by $\delta$ the angle that rotates the coordinates into the local coordinate system of the scaled natural stress variables, we have
\begin{align*}
    \delta =& \opatantwo\left(u_y, u_x\right)\, ,
\end{align*}
which, taking into account $C<0$ and thus $u_x,u_y<0$, yields
\begin{align}
    \delta =& \arctan \left(\frac{u_y}{u_x}\right) -\pi\\
        =& \arctan \left(\frac{\sin \theta}{3+\cos\theta}\right) -\pi\, .
\end{align}
Neglecting for the moment the additional $-\pi$ term, this is effectively the same result as for $\beta$ in~\eqref{eqn:sol_beta_core}. Using the fact that~\eqref{eqn:rot_identity_beta} not only describes the transformation of $\beta$ from Cartesian to polar coordinates, but can equally be applied to the transformation of $\beta$ from Cartesian to natural stress coordinates, we can introduce $\tilde{\tilde{\beta}}= \beta - 2\delta = 2\pi$, describing twice the angle that diagonalizes $\mathbf{\Psi}$, and thus $\mathbf{C}$, in the flow-aligned coordinates.

Consequently, \eqref{eqn:eigvaldecomppsi} yields
\begin{align}
    \begin{pmatrix} C_{uu} & C_{uw}\\ C_{uw} & C_{ww} \end{pmatrix}
        =& \begin{pmatrix} \exp\left(\frac{\optr\mathbf{\Psi}}{2} + \sqrt{D}\right) & 0 \\ 0 & \exp\left(\frac{\optr\mathbf{\Psi}}{2} - \sqrt{D}\right)\end{pmatrix}\, .
\end{align}
In other words, $\mathbf{C}$ is diagonal in scaled natural stress variables, and we have just shown that $\mu = 0$. This result goes beyond the existing discussions of the stick-slip singularity in the literature~\cite{evans2015stick,evans2019numerical}, which merely state $\mu = \mathcal{O}(r^0)$.

Comparing both sides, we can also determine $\lambda$ and $\nu$. To this end, notice that
\begin{align}
    \ln\,\seminorm{\mathbf{u}}^2 =& 2\ln\left(2\seminorm{C} r^{1/2} \sin\frac{\theta}{2}\right)
                + o(\theta)\, ,
\end{align}
which using~\eqref{eqn:sol_sqrtD_core} implies
\begin{align}\label{eqn:nat_stress_mu_core}
\begin{split}
    \ln\, \lambda =& \frac{\optr\mathbf{\Psi}}{2} + \sqrt{D} - \ln\,\seminorm{\mathbf{u}}^2\\
        =& -\frac{7}{8}\ln \left(2 r^{3/2} \sin\frac{\theta}{2}\sin\theta\right) + a_1\\&
            + \frac{1}{2} \ln\left(\frac{1-\alpha}{\alpha}\right) - \frac{5}{4}\gamma_0 - 2\ln\seminorm{C}\, .
\end{split}
\end{align}
Therefore, we recover exactly the relation $\lambda \propto (\tilde{\psi}/\seminorm{C})^{-7/8}$ to the stream function $\tilde{\psi}$, as known from the literature~\cite{evans2015stick,evans2019numerical}. Similarly, for $\nu$ we have
\begin{align}
\begin{split}
    \ln\nu =& \frac{7}{8}\ln \left(2 r^{3/2} \sin\frac{\theta}{2}\sin\theta\right) - a_1\\&
            + \frac{1}{2} \ln\left(\frac{1-\alpha}{\alpha}\right) + \frac{5}{4}\gamma_0 + 2\ln\seminorm{C}\, .
\end{split}
\end{align}
\end{rem}

\begin{rem}[Possibility of a different solution]
The existing literature on the asymptotic analysis of viscoelastic singularities, starting with~\cite{Renardy199491}, has centered very much around the idea that $\mathbf{u}\cdot\nabla \{\lambda,\mu,\nu\} = 0$ holds asymptotically, which justifies that the natural stress variables $\{\lambda,\mu,\nu\}$ must be functions of the stream function $\tilde{\psi}$ in the core region and then leads to the solutions above. It is therefore quite remarkable that our asymptotic analysis finds a second solution~\eqref{eqn:second_sol_utilde}, unknown to this previous analysis, which, although it does not match the viscometric behavior in the present situation, may well be of importance for other constitutive models or other settings, e.g., for studying bifurcations in the unsteady setting. It is thus a subject of further research to see whether there are situations in which this second solution materializes for a UCM-like fluid.
\end{rem}

\section{Slip boundary layer and transition zone}

Approaching the slip boundary, we first note that $\beta$ still has the same solution as in the core region, since the assumption $\coth\sqrt{D} \approx 1$ remains valid.

Furthermore, introducing $\vartheta = \pi -\theta$, we have the following small-angle approximations:
\begin{align}
    \omega_{r\theta} =& \frac{1}{4}\vartheta + \mathcal{O}(\vartheta^3)\\
    \epsilon_{rr} =& -\frac{1}{2} + \frac{7}{16}\vartheta^2 + \mathcal{O}(\vartheta^4)\\
    \epsilon_{r\theta} =& - \frac{1}{2}\vartheta + \mathcal{O}(\vartheta^3)\\
    \hat{u}_r =& -1 + \frac{7}{8}\vartheta^2 + \mathcal{O}(\vartheta^4)\\
    \hat{u}_\theta =& -\frac{3}{2} \vartheta + \mathcal{O}(\vartheta^3)\\
    \beta =& \vartheta + \mathcal{O}(\vartheta^3)\, .\label{eqn:beta_sol_slip}
 \end{align}
Using these small-angle approximations, we can simplify the differential equation~\eqref{eqn:asympsqrtDeqn} to
\begin{align}\label{eqn:sqrtd_eqn_slip}
    - \frac{\partial\sqrt{D}}{\partial \ln r} + \frac{3}{2} \frac{\partial\sqrt{D}}{\partial \ln \vartheta}
        =& -1 + 2 e^{\frac{1}{2}\ln r + \sqrt{D} + 2\gamma_0} + \mathcal{O}(\vartheta^2)\, .
\end{align}
Based on the solution from the core region, which serves as the initial solution that is advected into the slip region, we define
\begin{align}\label{eqn:def_v}
\begin{split}
    v(\ln r, \theta) =& \sqrt{D} + \frac{1}{4}\ln r + \gamma_0\\&\quad -\tilde{g}\left(-\frac{1}{4}\ln r + \ln(2\,\sin(\theta/2)) - \gamma_0\right)\\&\quad
                        + \frac{7}{8} \ln (\cos(\theta/2)) - o(\theta)\,,
\end{split}
\end{align}
which for small $\vartheta$ yields
\begin{align}\label{eqn:approx_v}
\begin{split}
    v(\ln r, \vartheta) =& \sqrt{D} + \frac{5}{16}\ln r + \frac{5}{4}\gamma_0\\&\quad
                        + \frac{7}{8} \ln (\vartheta)
                        +\frac{7}{8}\ln 2 - a_1\,.
\end{split}
\end{align}
As stated, $v$ is designed such that $v = 0$ holds everywhere except in close proximity to the slip boundary; formulated differently,
\begin{align}\label{eqn:approx_sqrtD_before_slip}
\begin{split}
    \sqrt{D} =&  - \frac{5}{16}\ln r - \frac{5}{4}\gamma_0
                        - \frac{7}{8} \ln (\vartheta)
                        -\frac{7}{8}\ln 2 + a_1
\end{split}
\end{align}
holds for small angles $\vartheta$ before the slip transition.

Considering the differential equation, notice that $(\partial_{\ln r} - \frac{3}{2}\partial_{\ln\vartheta}) (\frac{5}{16}\ln r + \frac{7}{8}\ln \vartheta) = -1$, which exactly cancels the constant $-1$ in~\eqref{eqn:sqrtd_eqn_slip}; this was the motivation for including the $\frac{7}{8} \ln (\cos(\theta/2))$ term already in the definition of $o$, namely to cancel the arising logarithmic singularity in $\vartheta$.

In particular, we can now write
\begin{align}
\begin{split}
    - \frac{\partial v}{\partial \ln r} + \frac{3}{2} \frac{\partial v}{\partial \ln \vartheta}
        =& 2 \exp \left(\frac{3}{16}\ln r - \frac{7}{8}\ln\vartheta\right.\\&\qquad\quad\left. + \frac{3}{4}\gamma_0 - \frac{7}{8}\ln 2 + a_1 + v\right)\, .
\end{split}
\end{align}

This motivates the introduction of the variable 
\begin{align}\label{eqn:def_y}
    y = \tfrac{3}{16}\ln r - \tfrac{7}{8}\ln\vartheta + \tfrac{3}{4}\gamma_0 - \tfrac{7}{8}\ln 2 + a_1\, ,
\end{align}
encoding the slip transition. Furthermore, we substitute
\begin{align}
    v(\ln r, \vartheta) =& \tilde{v}(y(\ln r, \ln\vartheta))\, .
\end{align}
It then follows that $\tilde{v}$ has to obey the differential equation
\begin{align}
    - \frac{3}{2} \frac{d\, \tilde{v}}{dy} =& 2\, e^y e^{\tilde{v}}\, .
\end{align}
Given that $\tilde{v}\to 0$ for $y\to-\infty$ is our initial condition, we obtain the solution
\begin{align}\label{eqn:sol_vtilde}
    \tilde{v}(y) =& -\ln \left(e^{y+2\ln 2 - \ln 3} + 1\right)\,,
\end{align}
which for large $y$ is approximately given by
\begin{align}
    \tilde{v}(y) \approx& - y - 2\ln 2 + \ln 3\, .
\end{align}
Based on~\eqref{eqn:approx_v}, this gives the slip solution
\begin{align}\label{eqn:slipbound_approx_sqrtD}
    \sqrt{D} \approx -\frac{1}{2}\ln r - 2\gamma_0 -2\ln 2 +\ln 3\, .
\end{align}

\begin{rem}[Lyapunov stability]\label{rem:lyapunov}
That this is the limiting solution for $\sqrt{D}$ in the vicinity of the slip boundary is not really a surprise. Using the substitution $w = \frac{1}{2}\ln r + \sqrt{D} + 2\gamma_0 + 2\ln 2 -\ln 3$, Eq.~\eqref{eqn:sqrtd_eqn_slip} amounts to the autonomous differential equation
\begin{align*}
    \frac{\partial w}{\partial \ln r} - \frac{3}{2} \frac{\partial w}{\partial \ln \vartheta}
        =& \frac{3}{2} \left(1-e^{w}\right) + \mathcal{O}(\vartheta^2)\, .
\end{align*}
The latter is Lyapunov stable at $w=0$, since $V(w) = \left(1-e^w\right)^2$ is a Lyapunov function.
\end{rem}

\begin{figure*}[h!]
    \centering
    \resizebox{\textwidth}{!}{%
        \input{tikz/stickslip_polar_graphical_abstract.tikz}
    }
    \caption{The piecewise solutions for $\operatorname{tr}\mathbf{\tilde{\Psi}}$, $\sqrt{D}$, and $\beta$ in the stick-slip case, shown in the scaled polar $(\ln r, \theta)$-plane together with the respective Newtonian velocity field. In the center plot, the left boundary corresponds to the singularity, the stick wall is at the bottom, and the slip wall is at the top. Close to the walls, the solutions are simplified by applying small-angle approximations. The transition lines mark where the Giesekus solution, for the given velocity field, switches from reaction-dominated (near the wall) to advection-dominated (away from the wall) behavior. The transition regions at both walls are highlighted by the respective log-log plots at the top and bottom, in which the transition lines become straight lines.}
    \label{fig:stickslip_polar_graphical_abstract}
\end{figure*}

\section{Full solution}

In total, we have now obtained a set of piecewise solutions to the log-conformation Giesekus equation at the stick-slip singularity, summarized in Fig.~\ref{fig:stickslip_polar_graphical_abstract}: in each of the three characteristic regions of the domain---along the stick wall, in the core, and along the slip wall---the degrees of freedom $\optr\mathbf{\Psi}$, $\sqrt{D}$, and $\beta$ take a different asymptotic form. In this section, we conclude the analysis by joining these piecewise results into a single composite solution for each of the three quantities:

\begin{itemize}
\item The trace of $\mathbf{\Psi}$ is determined once in the stick boundary-layer structure and remains constant thereafter, so that the full solution is simply given by
    \begin{align}\label{eqn:fullSoltrPsi}
        \optr\mathbf{\Psi}(\ln r,\theta) = \ln\left(\frac{1-\alpha}{\alpha}\right)\, .
    \end{align}
    It is the unique zero of the Giesekus relaxation term in the trace equation~\eqref{eqn:trPsi_eqn} in the limit $\cosh\sqrt{D} \gg 1$, independent of position, of the flow strength $C$, and of the relaxation time $\lambda$, and is solely set by the Giesekus mobility factor $\alpha$.
\item The derivation of $\sqrt{D}$ was already structured in such a way that $\sqrt{D}$ accumulates terms along the integration path. The only point requiring generalization concerns the arguments of $\tilde{g}$ and $\tilde{v}$. For $\tilde{g}$, we take into account that, for the transition to the core region, we chose in~\eqref{eqn:first_def_o} to lift the small-angle condition by replacing $\ln\theta$ with $\ln(2\sin(\theta/2))$, which motivates the following generalization of the argument of $\tilde{g}$:
    \begin{align}\label{eqn:X_transition_var_whole_domain}
    X(\ln r,\theta) = -\frac{1}{4}\ln r + \ln\left(2\sin(\theta/2)\right) - \gamma_0\, .
    \end{align}
    For the argument of $\tilde{v}$, we want the full solution to account for the stability of the ODE (cf.\ Remark~\ref{rem:lyapunov}), which motivates the generalization of the argument to
    \begin{align}\label{eqn:Y_transition_var_whole_domain}
    Y(\ln r,\theta) = \frac{1}{4}\ln r + \gamma_0 + \tilde{g}(X) + o(\theta) - \frac{7}{8}\ln(\cos (\theta/2))\, .
    \end{align}
    The full solution for $\sqrt{D}$ assembles the stick, core, and slip contributions into
    \begin{align}\label{eqn:fullSolsqrtD}
    \begin{split}
        \sqrt{D}(\ln r,\theta) =& - \frac{1}{4}\ln r - \gamma_0 + \tilde{g}(X) + o(\theta)\\
        &\quad - \frac{7}{8} \ln (\cos(\theta/2)) + \tilde{v}(Y)\, ,
    \end{split}
    \end{align}
    with the constant $\gamma_0$ from Eq.~\eqref{eqn:gamma_0}, the argument $X$
    encoding the stick transition, the solution $\tilde{g}$ of the boundary-layer ODE system~\eqref{eqn:ode_gtilde} with its log-rational representation~\eqref{eqn:pade_gtilde} and the coefficients in Tab.~\ref{tab:pade_coeffs}, the solution $o$ of the core equation~\eqref{eqn:core_o_eqn} given by~\eqref{eqn:sol_o_core}, and $\tilde{v}$ given in closed form by Eq.~\eqref{eqn:sol_vtilde} with the argument $Y$
    encoding the slip transition.
\item For the full solution for $\beta$, first note that the results were derived in terms of $\tilde{u} = \tan\frac{\beta}{2}$, such that it is natural to match the stick and core solutions at that level.
    Starting with the stick boundary, where~\eqref{eqn:approx_tildeu_stick} determines $\tilde{u}$, we lift the small-angle approximation in the same way as we did for $X$ and $\sqrt{D}$ by replacing $\ln \theta$ with $\ln\left(2\sin\frac{\theta}{2}\right)$, which yields
    \begin{align*}
        \tilde{u} =& 2\sin\frac{\theta}{2}\left(\frac{3}{8} - e^{-X+\tilde{g}(X)} - \frac{1}{2}\frac{d\tilde{g}}{dx}\right)\, .
    \end{align*}
    Subtracting $\frac{1}{2}\sin\frac{\theta}{2}$ to account for the fact that the stick and core solutions match with $\partial_\theta \tilde{u} = 1/4$ for small $\theta$, and then adding the core solution~\eqref{eqn:sol_utilde_core}, yields
    \begin{align}\label{eqn:fullSolbeta}
    \begin{split}
        \beta(\ln r,\theta) =&\; 2\arctan\left(\sin (\theta/2)\left(\frac{1}{4} - \tilde{g}'(X)\right) - e^{\frac{1}{4}\ln r + \gamma_0 + \tilde{g}(X)}\right.\\&\left.\qquad\qquad\qquad + \frac{\sin \theta}{3+\cos\theta}\right)\, .
    \end{split}
    \end{align}
    The derivative $\tilde{g}'$ is given in closed form by differentiating the rational approximation~\eqref{eqn:pade_gtilde} with respect to its scalar argument:
    \begin{align}\label{eqn:gtilde_derivative}
    \tilde{g}'(X) = \frac{e^{-X/8}}{4}\left(\frac{q'(e^{-X/8})}{q(e^{-X/8})} - \frac{p'(e^{-X/8})}{p(e^{-X/8})}\right)\, .
    \end{align}
\end{itemize}

\begin{rem}[Renormalization Group]
It should be noted that $\tfrac{1}{4} \ln r$ always appears in our asymptotic solution in combination with $\gamma_0$, which implies that our asymptotic solution fulfills the following Renormalization Group~(RG) equation:
\begin{align}
    2 \frac{\partial \mathbf{\Psi}}{\partial \ln r} + \frac{\partial \mathbf{\Psi}}{\partial \ln \lambda} =& 0\, .
\end{align}
That such a property is desirable for the asymptotic solution could already be anticipated from~\eqref{eqn:polar_logconf_rescaled}.

Although we have not made any explicit use of the Renormalization Group here, we note that it is a very important concept in quantum field theory~\cite{ZinnJustin2002} and has also found application in the asymptotic solution of differential equations~\cite{chen1994renormalization}.
\end{rem}

\begin{rem}[Tensor component reconstruction]
    The components of the log-conf tensor can be reconstructed by inserting Eqs.~\eqref{eqn:fullSoltrPsi}, \eqref{eqn:fullSolsqrtD}, and~\eqref{eqn:fullSolbeta} into Eq.~\eqref{eqn:eigvaldecomppsi}. A representation of $\mathbf{\Psi}$ in the Cartesian frame follows by evaluating $\mathbf{Q}$ with the Cartesian orientation angle $\beta$; correspondingly, a representation in the polar frame follows via the polar orientation angle $\tilde{\beta} = \beta - 2\theta$.
\end{rem}

\begin{figure*}[t!]
    \centering
    \begin{minipage}[t]{0.47\textwidth}
        \centering
        \includegraphics[width=\linewidth]{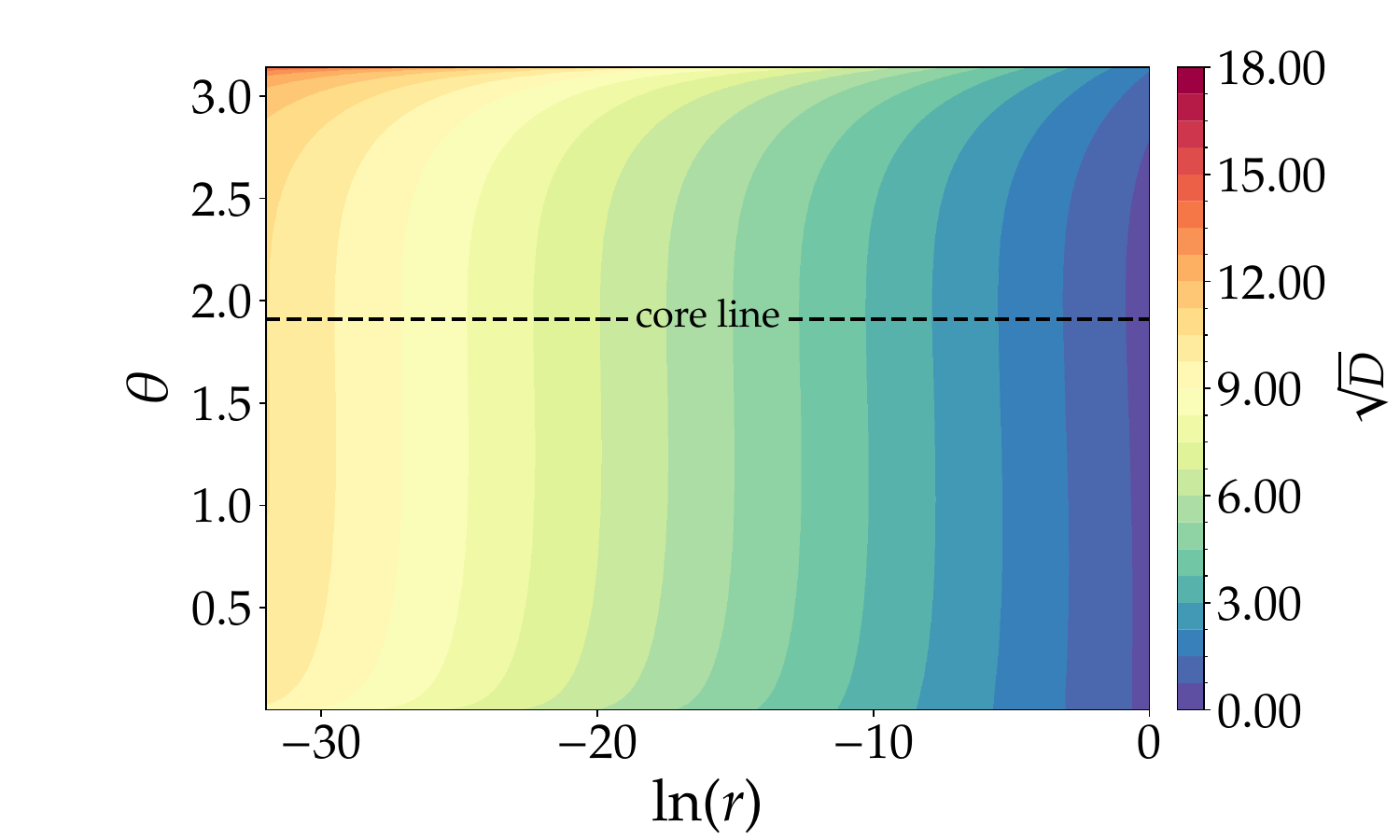}
    \end{minipage}
    \hspace{0.02\textwidth}
    \begin{minipage}[t]{0.47\textwidth}
        \centering
        \includegraphics[width=\linewidth]{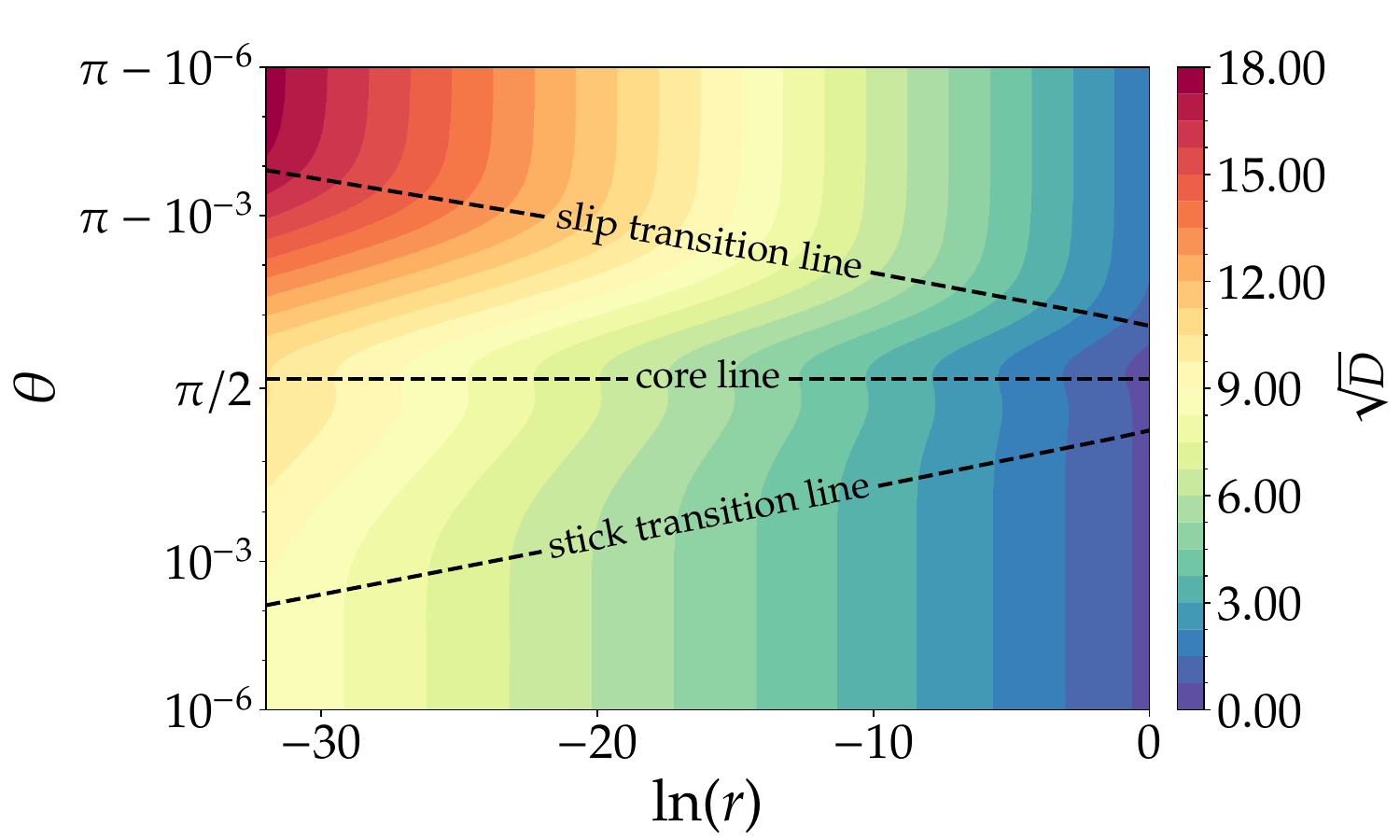}
    \end{minipage}
    \caption{Evaluation of $\sqrt{D}$ according to Eq.~\eqref{eqn:fullSolsqrtD} on the rectangular domain $(\ln r, \theta) \in [-32,0] \times [10^{-6}, \pi-10^{-6}]$, with the minimum distance to the singularity therefore being $r_{\mathrm{min}} = e^{-32} \approx 1.266\times10^{-14}$. The constant $\gamma_0 \approx -0.949$, needed for the evaluation of $\sqrt{D}$, is specified via $\alpha = 0.1$, $\lambda = 1$, and $C = -1$. The left plot shows the result in the standard $(\ln r, \theta)$ coordinates. The right plot shows the same data with a logarithmic scaling of the angular axis from $\pi/2$ toward the top and bottom boundaries, in order to highlight the stick and slip boundary layers and transition regions. It should be pointed out that the result shown only holds asymptotically for $\ln r \to -\infty$, and proper agreement with the real physical behavior of a Giesekus fluid should therefore only be expected for small $r$.}
    \label{fig:sqrtD_evaluation}
\end{figure*}

Since the asymptotic analysis predicts a constant trace and an orientation angle $\beta$ (or $\tilde{\beta}$) that is asymptotically independent of $\ln r$, the singular behavior of the log-conf tensor is entirely encoded in $\sqrt{D}$. We therefore conclude the analysis by examining the analytical $\sqrt{D}$ solution~\eqref{eqn:fullSolsqrtD}. Fig.~\ref{fig:sqrtD_evaluation} shows $\sqrt{D}$ according to Eq.~\eqref{eqn:fullSolsqrtD} in the rectangular $(\ln r, \theta)$ domain.
\begin{figure*}[ht!]
    \centering
    \begin{minipage}[t]{0.47\textwidth}
        \centering
        \includegraphics[width=\linewidth]{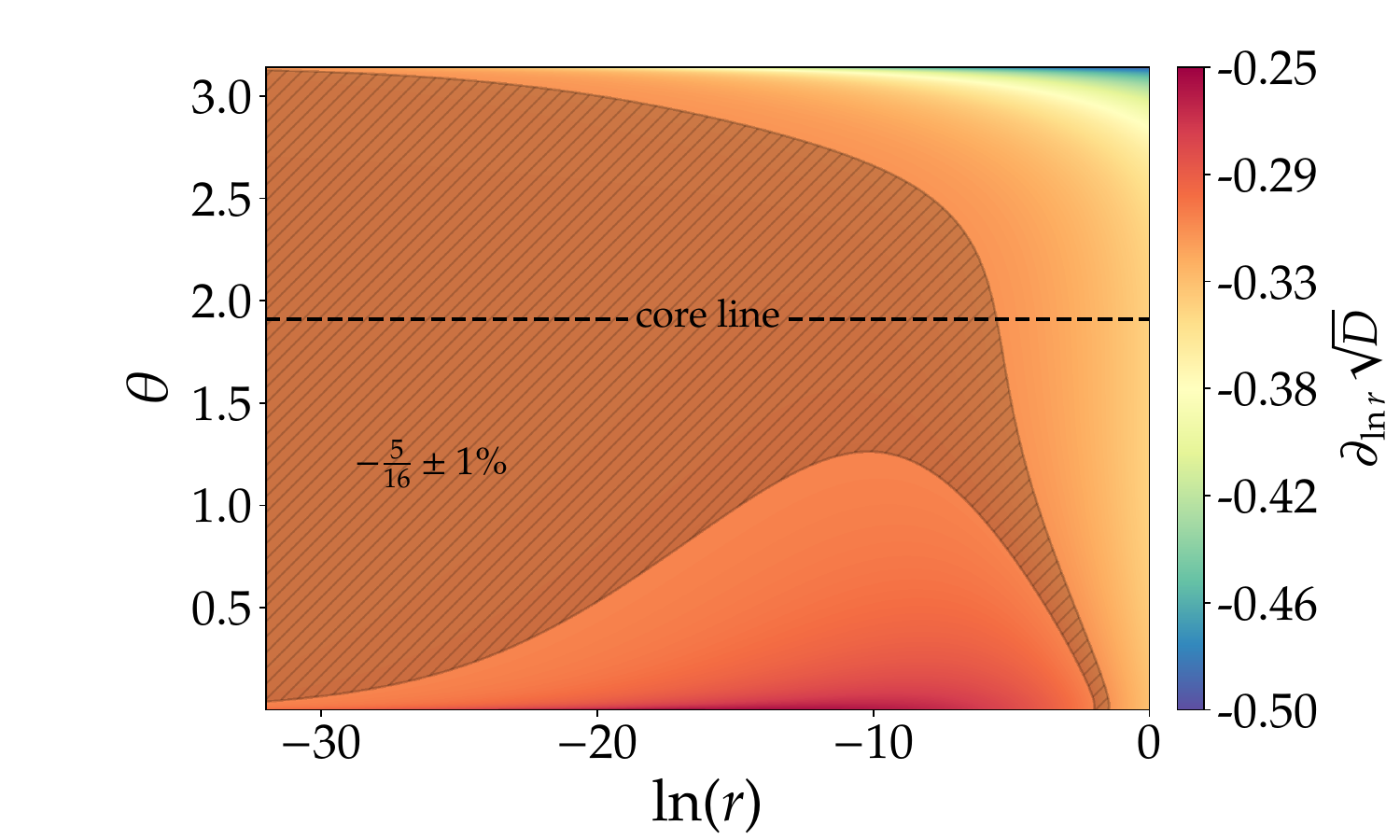}
    \end{minipage}
    \hspace{0.02\textwidth}
    \begin{minipage}[t]{0.47\textwidth}
        \centering
        \includegraphics[width=\linewidth]{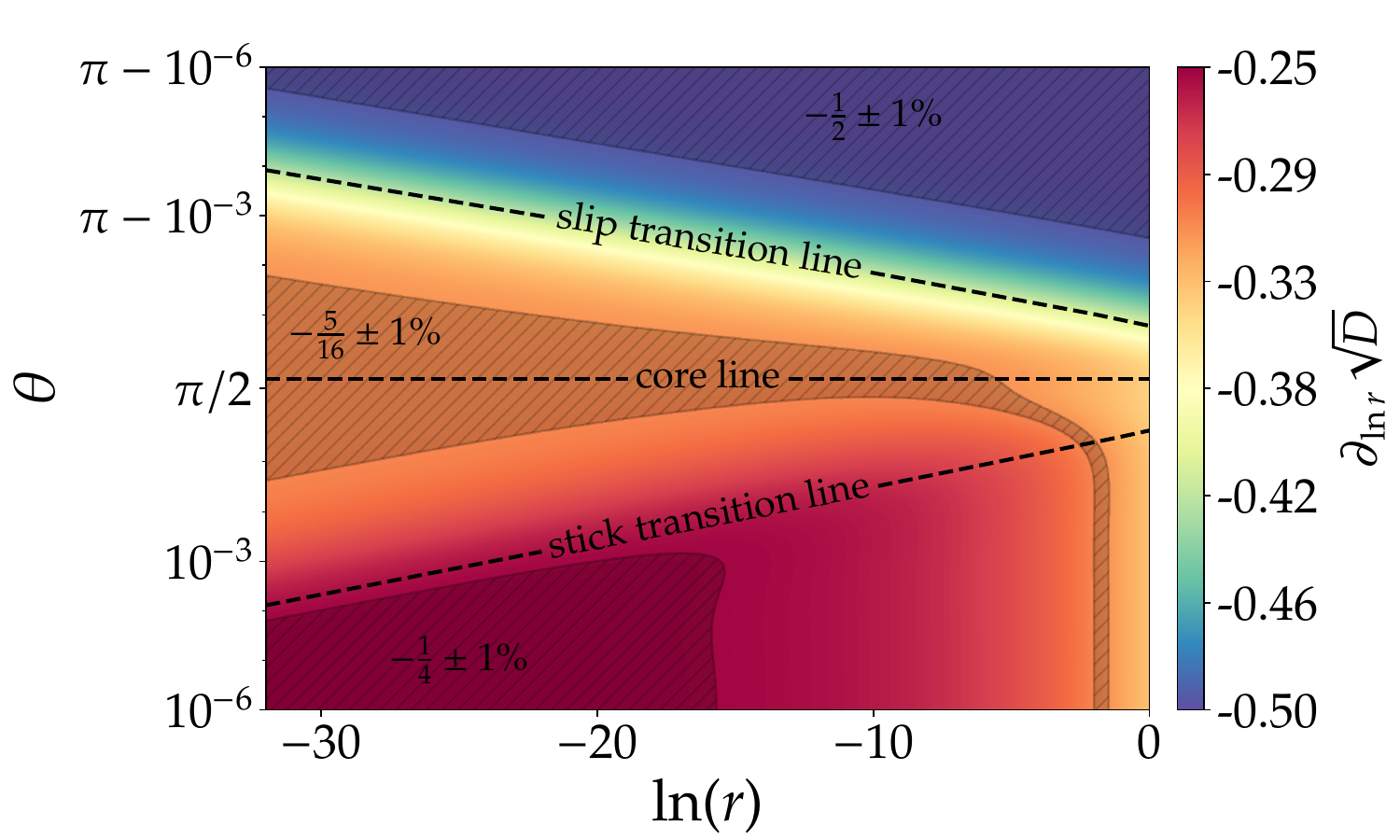}
    \end{minipage}
    \caption{Evaluation of $\partial_{\ln r}\sqrt{D}$ according to Eq.~\eqref{eqn:dsqrtDdlnr} under the same conditions as in Fig.~\ref{fig:sqrtD_evaluation}. Shaded areas mark where the solution lies within $\pm1\%$ of the characteristic slopes of $-1/4$ inside the stick boundary layer, $-5/16$ in the core region, and $-1/2$ inside the slip boundary layer. As before, the boundary-layer behavior only becomes visible in the scaled plot on the right.}
    \label{fig:sqrtD_dlnr_evaluation}
\end{figure*}
Overall, the value of $\sqrt{D}$ increases monotonically everywhere with decreasing $\ln r$. For a fixed and sufficiently small $\ln r$, the lowest values occur directly at the stick wall. As $\theta$ increases, the right plot shows the constant behavior of $\sqrt{D}$ inside the stick boundary layer, where the solution sits at the value $-\tfrac{1}{4} \ln r - \gamma_0$ according to Eq.~\eqref{eqn:sqrtD_before_stick_trans}. After crossing the stick transition line at $X(\ln r, \theta) = 0$, Eq.~\eqref{eqn:X_transition_var_whole_domain}, the core region is entered. After a short stick transition phase, the value of $\sqrt{D}$ stays nearly flat inside the core region until the transition into the slip boundary layer sets in and $\sqrt{D}$ increases significantly more strongly. After crossing the slip transition line at $Y(\ln r, \theta) = 0$, Eq.~\eqref{eqn:Y_transition_var_whole_domain}, $\sqrt{D}$ reaches its final constant value $-\tfrac{1}{2} \ln r - 2\gamma_0 -2 \ln 2 + \ln 3$ according to~\eqref{eqn:slipbound_approx_sqrtD}.

The behavior of $\sqrt{D}$ over $\theta$ is of course closely connected to the three characteristic slopes of $\sqrt{D}$ toward the singularity at the stick wall, along the core line, and at the slip wall. By applying the chain rule to the solution~\eqref{eqn:fullSolsqrtD} with respect to $\ln r$, we obtain the radial derivative as
\begin{align}\label{eqn:dsqrtDdlnr}
\partial_{\ln r}\sqrt{D} = -\tfrac{1}{4} + \partial_{\ln r} X \, \tilde{g}'(X) + \partial_{\ln r} Y \, \tilde{v}'(Y)\, .
\end{align}
The partial derivatives of $X$ and $Y$ are given by
\begin{align}
\partial_{\ln r} X &= -\tfrac{1}{4} \label{eqn:Xderivative} \\
\partial_{\ln r} Y &= \tfrac{1}{4} - \tfrac{1}{4}\,\tilde{g}'(X)\, . \label{eqn:Yderivative}
\end{align}
Furthermore, $\tilde{g}'(X)$ is given by~\eqref{eqn:gtilde_derivative}, and
\begin{align}\label{eqn:vtildederivative}
\tilde{v}'(Y) = -\big(1 + \tfrac{3}{4}e^{-Y}\big)^{-1}\, ,
\end{align}
follows from differentiating Eq.~\eqref{eqn:sol_vtilde}. This closed-form expression for the radial derivative of $\sqrt{D}$ is plotted in Fig.~\ref{fig:sqrtD_dlnr_evaluation}.
The shaded regions in Fig.~\ref{fig:sqrtD_dlnr_evaluation} mark where our theory predicts the radial slope to lie within $1\%$ of one of the three asymptotic values. The plots clearly identify the characteristic slopes of the Giesekus fluid with $\partial_{\ln r}\sqrt{D} = -1/4$ inside the stick boundary layer, $-5/16$ in the core region, and $-1/2$ in the slip boundary layer. In particular, each of these slopes is found strictly within its respective region of the stick-slip domain, separated by thin transition zones around the respective transition lines.

\begin{figure*}[t!]
    \centering
    \begin{minipage}[t]{0.47\textwidth}
        \centering
        \includegraphics[width=\linewidth]{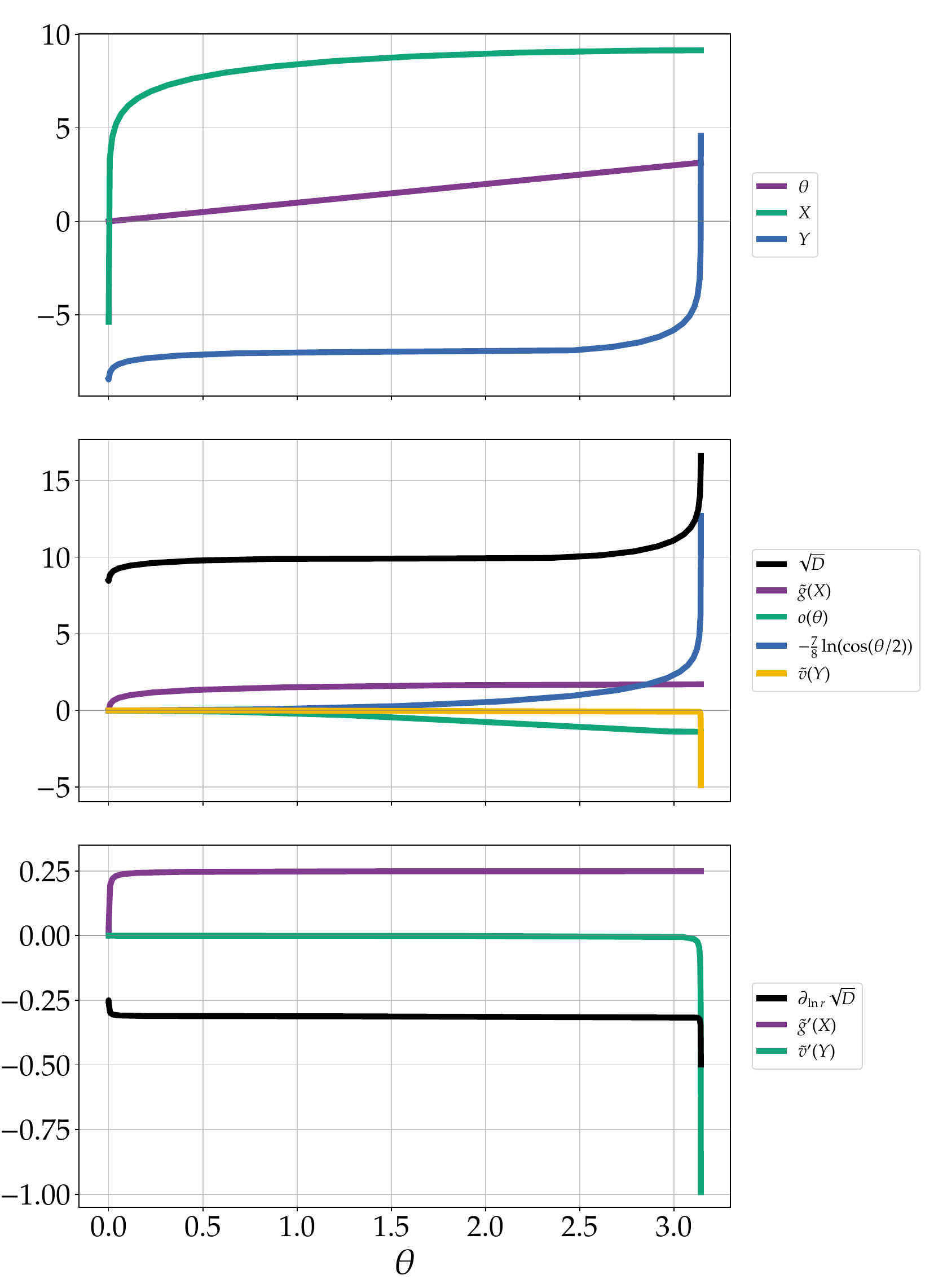}
    \end{minipage}
    \hspace{0.02\textwidth}
    \begin{minipage}[t]{0.47\textwidth}
        \centering
        \includegraphics[width=\linewidth]{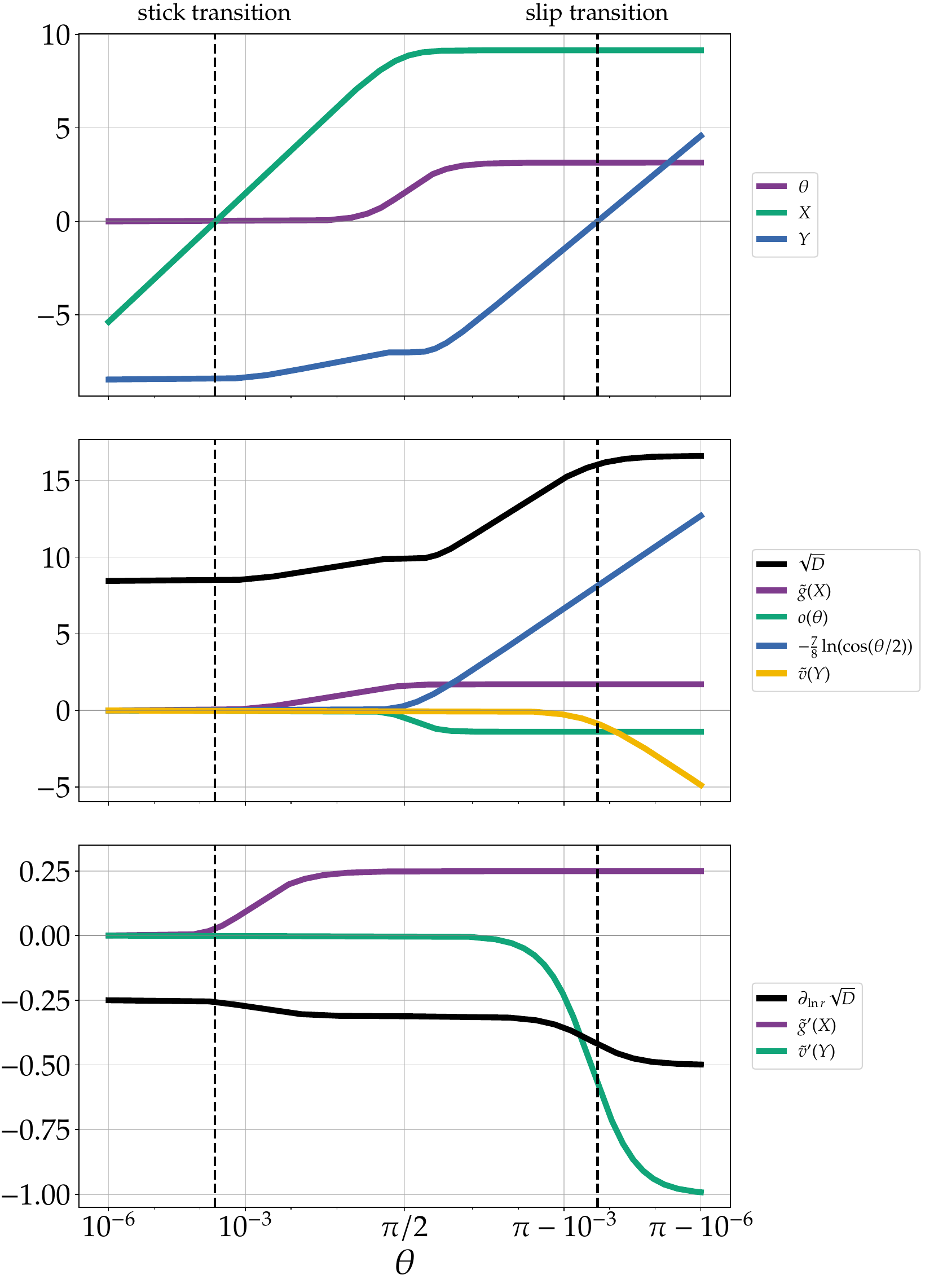}
    \end{minipage}
    \caption{The decomposition of $\sqrt{D}$ and $\partial_{\ln r}\sqrt{D}$, according to Eqs.~\eqref{eqn:fullSolsqrtD} and~\eqref{eqn:dsqrtDdlnr}, respectively, into their $\theta$-dependent components at $\ln r = -30$. Again, the expressions are evaluated under the same conditions as in Fig.~\ref{fig:sqrtD_evaluation}. The left and right plots show the same data with different scalings of the $\theta$ axis. The dashed lines in the right plot indicate the transition lines, given by the $\theta$ values at which $X(-30,\theta) = 0$ and $Y(-30,\theta) = 0$.}
    \label{fig:sqrtD_cmpts}
\end{figure*}

At fixed $\ln r$, the $\theta$-dependence of $\sqrt{D}$ in Eq.~\eqref{eqn:fullSolsqrtD} is carried by $\tilde{g}(X)$, $o(\theta)$, $-\frac{7}{8} \ln (\cos(\theta/2))$, and $\tilde{v}(Y)$. Fig.~\ref{fig:sqrtD_cmpts} shows their contributions to $\sqrt{D}$ (and $\partial_{\ln r}\sqrt{D}$) over $\theta$ and thereby illustrates the construction behind~\eqref{eqn:fullSolsqrtD}.
The top panel shows the arguments entering~\eqref{eqn:fullSolsqrtD}, namely $\theta$, $X$, and $Y$. As $\theta$ increases from $0$ to $\pi$, $X$ encodes the stick transition by switching from negative to positive at a specific $\theta$ value and leveling off logarithmically toward the slip wall at $\theta = \pi$. $Y$ also starts negative. It is lifted by $\tilde{g}$ across the stick transition but remains negative, stays almost constant through the core, and increases sharply only in the vicinity of the slip wall, where it encodes the slip transition by changing its sign.

The center panel shows the decomposition of $\sqrt{D}$: in the stick boundary layer, $\tilde{g}$, $o$, $- \frac{7}{8} \ln (\cos(\theta/2))$, and $\tilde{v}$ are all almost zero, such that $\sqrt{D}$ stays at a constant value. The boundary-layer term $\tilde{g}$ is then activated as its argument $X$ crosses $0$ beyond the stick transition, follows its asymptote~\eqref{eqn:approxgtilde}, and levels off once $X$ saturates logarithmically toward the slip side. Meanwhile, $o(\theta)$ contributes monotonically from $o(0) = 0$ down to its slip-wall value $o(\pi) = -2\ln 2 \approx -1.386$. The term $-\tfrac{7}{8}\ln(\cos(\theta/2))$ starts near zero in the stick region; as $\theta$ moves through the core region and approaches $\pi$, it diverges toward $+\infty$. This is countered by the last term, $\tilde{v}$, which is approximately $0$ before the slip transition, where $Y \ll 0$. Once $Y$ crosses $0$, $\tilde{v}(Y) \approx -Y - 2\ln 2 + \ln 3$ cancels the divergence of $-\tfrac{7}{8}\ln(\cos(\theta/2))$, and $\sqrt{D}$ saturates at its slip plateau of Eq.~\eqref{eqn:slipbound_approx_sqrtD}.

The bottom panel assembles the radial derivative from Eq.~\eqref{eqn:dsqrtDdlnr}, with both derivative factors evaluated along $X$ and $Y$. At the stick wall, $\tilde{g}'(X) = 0$ and $\tilde{v}'(Y) = 0$ for $\theta \to 0$, hence $\partial_{\ln r} Y = \tfrac{1}{4}$ and $\partial_{\ln r}\sqrt{D} = -\tfrac{1}{4}$. Across the stick transition, $\tilde{g}'(X)$ rises to $\tfrac{1}{4}$, so $\partial_{\ln r} Y$ drops to $\tfrac{3}{16}$ and the slope settles at the core value $-\tfrac{1}{2} + \tfrac{3}{16} = -\tfrac{5}{16}$, still unaffected by $\tilde{v}'$. Only within the slip layer does $\tilde{v}'(Y)$ fall from $0$ to $-1$, suppressing the $\partial_{\ln r} Y$ contribution entirely, so that the slope saturates at $-\tfrac{1}{2}$.

\section{Conclusion}
The steady planar stick-slip benchmark with a prescribed Newtonian velocity field was analyzed for a Giesekus fluid in the log-conformation formulation, using polar coordinates with $\ln r$ as the radial coordinate. The analysis was carried out in the variables $(\optr\mathbf{\Psi}, \sqrt{D}, \beta)$---the trace, the deviatoric magnitude, and the orientation angle---which parametrize the symmetric $2\times 2$ log-conf tensor $\mathbf{\Psi}$ through its eigenvalue decomposition~\eqref{eqn:eigvaldecomppsi}. Composite asymptotic solutions for all three quantities were derived in Eqs.~\eqref{eqn:fullSoltrPsi}, \eqref{eqn:fullSolsqrtD}, and~\eqref{eqn:fullSolbeta}. The solutions are formulated in the $(\ln r,\theta)$-plane, hold asymptotically for $\ln r \to -\infty$ uniformly in $\theta\in[0,\pi]$, and yield the full log-conf tensor in Cartesian or polar components via Eq.~\eqref{eqn:eigvaldecomppsi}.

The construction also identifies the stick and slip transition lines as the zero sets of $X$ and $Y$ in Eqs.~\eqref{eqn:X_transition_var_whole_domain} and~\eqref{eqn:Y_transition_var_whole_domain}, along which the constitutive behavior changes from reaction-dominated in the wall layers to advection-dominated in the core: the stick line separates the stick layer from the core, and the slip line separates the core from the slip layer. They scale as $\theta \propto r^{1/4}$ and $\pi-\theta \propto r^{3/14}$, respectively. In the three regions separated by these lines, $\sqrt{D}$ grows linearly in $\ln r$ with the characteristic slopes $\partial_{\ln r}\sqrt D = -1/4$ at the stick wall, $-5/16$ in the core, and $-1/2$ at the slip wall, consistent with the results of Evans~\cite{evans2015stick}.

Beyond the consistency with the known slopes, the composite solutions open up a new route for benchmarking numerical Giesekus solvers. Since Eqs.~\eqref{eqn:fullSoltrPsi}, \eqref{eqn:fullSolsqrtD}, and~\eqref{eqn:fullSolbeta} hold uniformly in $\theta$, numerical results can be compared with the asymptotic solution on the whole domain, either in the $(\optr\mathbf{\Psi},\sqrt{D},\beta)$ parametrization or, as is more customary for numerical codes, directly in the tensor components reconstructed via Eq.~\eqref{eqn:eigvaldecomppsi}. The $\ln r$-scaled polar log-conf equation~\eqref{eqn:polar_logconf_rescaled} lends itself to this purpose: it is posed on the rectangular $(\ln r,\theta)$ domain, where meshing is straightforward and the singularity can be approached arbitrarily closely by extending the mesh toward $\ln r\to-\infty$. With prescribed Newtonian kinematics, only the constitutive equation has to be solved, which decouples the log-conf equation from the momentum and mass balances and thereby provides a clean, isolated test environment for viscoelastic codes. The asymptotic solutions could equally serve as a reference for simulations in the physical plane of Fig.~\ref{fig:stickslip_kartesian_illu}.

\section*{Acknowledgments}
We would like to thank in particular Nick Jaensson, Felix Terhag, Fabian Hoppe, and many others for encouraging us to finish this work.

Part of this work was supported by the Federal Ministry for Economic Affairs and Energy (BMWE) on the basis of a decision by the German Bundestag.

\section*{Declaration of AI assistance}
The authors used Claude (Anthropic) during the preparation of this manuscript for the following specific, auxiliary tasks: (i) generating TikZ code to produce illustrative graphics; (ii) drafting SymPy scripts to perform independent arithmetic and symbolic consistency checks; (iii) verifying the algebraic manipulation of final formulas; and (iv) copy-editing and linguistic polishing of the prose. All mathematical content---including the research questions, theorems, proofs, derivations, and the initial manuscript text---was conceived, developed, and written entirely by the authors. Any AI-generated code or suggestions were manually reviewed, validated, and, where necessary, corrected by the authors. The AI tool was not used to generate research ideas, mathematical arguments, or scientific conclusions. The authors assume full responsibility for all content.

\bibliographystyle{elsarticle-num}
\bibliography{refs}
\end{document}